\documentclass[useAMS, usenatbib]{mn2e} 
\usepackage{times}
\usepackage{url}
\usepackage{amsmath}
\usepackage{amsfonts}
\usepackage{amssymb}
\usepackage{graphicx}
\usepackage{subfig}
\usepackage{float}
\usepackage{caption}
\usepackage{multirow}
\usepackage{color}
\usepackage{xspace}
\usepackage[breaklinks, colorlinks, citecolor=blue]{hyperref}
\usepackage{txfonts}
\usepackage{algpseudocode}
\usepackage{algorithmicx}
\usepackage{algorithm}

\newcommand{\half}{\frac{1}{2}}

\newcommand{\ie}{\textit{i.e.,~}}
\newcommand{\eg}{\textit{e.g.,~}}
\newcommand{\equref}[1]{Eq.~(\ref{#1})}
\newcommand{\equrefs}[2]{Eqs.~(\ref{#1}) and~(\ref{#2})}
\newcommand{\equrefss}[3]{Eqs.~(\ref{#1}), (\ref{#2}) and~(\ref{#3})}
\newcommand{\figref}[1]{Fig.~\ref{#1}}
\newcommand{\secref}[1]{Sec.~\ref{#1}}
\newcommand{\ang}{{\bmath{n}}}

\newcommand{\spinpm}{\ensuremath{\eth_{\pm}}}

\newcommand{\equ}[1]{\begin{equation}#1\end{equation}}
\newcommand{\eqn}[1]{\begin{eqnarray}#1\end{eqnarray}}

\newcommand{\sshtcode}{{\sc ssht}}
\newcommand{\sothreecode}{{\sc so3}}
\newcommand{\stwoletcode}{{\sc s2let}}
\newcommand{\ebsepcode}{{\sc ebsep}}
\newcommand{\healpix}{{\sc healpix}}
\newcommand{\fftwcode}{{\sc fftw}}

\newcommand{\wav}{\ensuremath{\Psi^j}}

\newcommand{\Q}{\(Q\)\xspace}
\newcommand{\U}{\(U\)\xspace}
\newcommand{\E}{\(E\)\xspace}
\newcommand{\B}{\(B\)\xspace}

\usepackage{color}

\pubyear{2016}
\def\LaTeX{L\kern-.36em\raise.3ex\hbox{a}\kern-.15em
    T\kern-.1667em\lower.7ex\hbox{E}\kern-.125emX}

\title[\E-\B separation with spin directional wavelets]
{Wavelet reconstruction of \E and \B modes for CMB polarisation and cosmic shear analyses}

\author[Leistedt et al.]
  {Boris~Leistedt,$^{1,2}$ Jason~D.~McEwen,$^3$ Martin~B\"uttner,$^2$ and Hiranya~V.~Peiris$^2$\\
  $^1$Center for Cosmology and Particle Physics, Department of Physics, New York University, New York, NY 10003, USA\\
  $^2$Department of Physics and Astronomy, University College London, London WC1E 6BT, U.K \\
  $^3$Mullard Space Science Laboratory (MSSL), University College London (UCL), Surrey RH5 6NT, UK \\
  Email: boris.leistedt@nyu.edu}

\begin{document}

\maketitle 

\begin{abstract}
We present new methods for mapping the curl-free (\E-mode) and divergence-free (\B-mode) components of spin 2 signals using spin directional wavelets. Our methods are equally applicable to measurements of the polarisation of the cosmic microwave background (CMB) and the shear of galaxy shapes due to weak gravitational lensing. We derive pseudo and pure wavelet estimators, where \E-\B mixing arising due to incomplete sky coverage is suppressed in wavelet space using scale- and orientation-dependent masking and weighting schemes.  In the case of the pure estimator, ambiguous modes (which have vanishing curl and divergence simultaneously on the incomplete sky) are also cancelled. On simulations, we demonstrate the improvement (\ie reduction in leakage) provided by our wavelet space estimators over standard harmonic space approaches. Our new methods can be directly interfaced in a coherent and computationally-efficient manner with component separation or feature extraction techniques that also exploit wavelets. 
\end{abstract}
\begin{keywords}
Wavelets, cosmic microwave background polarisation, cosmic shear, observational cosmology.
\end{keywords}

\section{Introduction}

The cosmic microwave background (CMB) polarisation \citep[see \eg][]{1997PhRvL..78.2054S, spergel:1997, 2009astro2010S..67D} and cosmic shear \citep[see \eg][]{Albrecht:2006um, peebles:2003, Heavens:2006uk, Peacock:2006kj, Weinberg:2012es}, measured via distortions of galaxy shapes due to gravitational lensing, are two of the key cosmological observables targeted by the next generation of astronomical surveys, offering insights into the physics of both the early and evolved Universe. These observables have a key property in common: they are both spin 2 signals.  Consequently, these observables behave in a simple manner under local rotations of the tangent plane but are not invariant with respect to changes of the local coordinate system.  
They are, however, invariant under local rotations of $180$ degrees in the tangent plane.  Theoretical models usually make predictions for the global curl-free (\E-mode) and divergence-free (\B-mode) parts of these signals, which are scalar and pseudo scalar quantities, respectively, giving rise to the spin 2 observables (the CMB quantity formed from the Stokes parameters $Q\pm iU$ and the complex shear $\gamma$).
Extracting \E and \B modes from CMB polarisation or cosmic shear observations is challenging but essential for confronting data with theoretical predictions.

For both observables, \E and \B mode estimators have been developed, but these are mostly focused on the estimation of summary statistics such as \E-\B mode angular power spectra or correlation functions \citep[\eg][]{lewis:2002a, Hivon:2001jp, Chon:2003gx, Schneider:2010pm, Becker:2014kna}.
In the CMB case, a significant body of work focuses on understanding the mixing that occurs between \E and \B modes.
Such leakage occurs when analysing spin 2 signals on the incomplete sky, where the \E-\B mode decomposition is not unique. This is relevant to all CMB and cosmic shear experiments since most observations are made on partial regions of the sky and even full sky data sets are subject to extra masking, \eg to remove foreground contamination. 

On the cut sky, \E and \B modes can no longer be unambiguously identified; they fall in two categories: {\it pure} and {\it ambiguous} modes. Pure \E (\B) modes are orthogonal to all \B (\E) modes in the sky region of interest, regardless of the particular realisation of the spin 2 field.\footnote{Thus, the sets of pure and ambiguous modes are uniquely defined by the region of the sphere to be analysed.} Ambiguous modes have vanishing curl and divergence simultaneously on the sky region of interest. Hence when splitting the spin signal into \E and \B these ambiguous modes can go into either component. Because ambiguous modes can be assigned to \E or \B, they contribute to the estimated \E and \B maps and power spectra. 
For the CMB, as the power spectrum of \E modes is much larger than that of \B modes, these ambiguous modes significantly increase the variance of the estimated \B-mode power spectrum.
This jeopardises the detection of inflationary gravitational waves since their amplitude is small relative to this \E-to-\B leakage.
 
Pseudo-power spectrum estimators including only pure \E and \B modes have been developed \citep[\eg][]{Lewis:2003an, bunn:2003, Smith2006pureEB, SmithZaldarriaga2007pureEB,  Grain2012crosspure, Ferte2014pureEB} and are a powerful, computationally-efficient alternative to maximum likelihood estimators.\footnote{The latter implicitly performs the optimal \E-\B decomposition on the cut sky and delivers minimum-variance power spectrum estimates; however it is computationally prohibitive for current data sets.}
The central idea behind pure mode estimators is to apply a suitable weight function to apodise the mask such that the ambiguous modes (thus, the \E-\B leakage) can be removed explicitly.
The amount of cancellation is determined by the weighting scheme used to apodise the mask. 
This process can be optimised when the mask and noise properties of the data set are known, and even carried out as a function of power spectrum band powers \citep[\eg][]{SmithZaldarriaga2007pureEB}.
In the case of the CMB, these weights must be optimised to account for the fact that the power spectrum of \E modes is much larger than that of \B modes.
This has the effect of removing most of the ambiguous modes from the \B-mode maps or power spectrum.
Even though some cosmological information is lost in this process, it also removes the large variance induced by the ambiguous modes. 
Both effects are self-consistently accounted for in pure estimators, which are unbiased and close to optimal in the case of the CMB \B modes.
For this reason they are critical in the search for inflationary gravitational waves.

While power spectra are powerful summary statistics, mapping the \E-\B modes on the sky is also of great interest.
For instance, distinctive patterns in \E-\B modes are expected around hot and cold spots in the CMB, and have been shown to be useful cosmological probes \citep[\eg][]{Ade:2013nlj, Ade:2015hxq}.
Similarly, massive structures such as galaxy clusters (or voids, which are underdense analogues of clusters) have distinctive effects on galaxy shapes and therefore the cosmic shear \E-\B modes \citep[\eg][]{Pujol:2016lfe, Gruen:2015jhr}.
Exploiting these new observables requires accurate methods for mapping the \E-\B modes in the presence of complicated sky cuts and noise, which cause nontrivial \E-\B mixing.

Pure mode, harmonic space estimation techniques could support map making \citep[as highlighted in \eg][]{Grain2012crosspure, Ferte2014pureEB} but to date have not been used for this purpose. 
Also, optimising the weighting schemes as a function of scale or orientation on the sky has not been investigated. 
Other techniques for CMB \E and \B-mode reconstruction in pixel space \citep[\eg][which use finite differencing]{Bowyer:2010yp, Bowyer:2011sg} or harmonic space \citep[\eg][involving fine-tuned apodisation parameters]{kim:2011} do not support \E-\B leakage cancellation, therefore yielding maps with greater leakage.
Similar efforts in the context of cosmic shear have focused on directly reconstructing the underlying convergence field using flat sky estimators \citep[\eg][]{Kaiser:1992ps} and lead to deep 2D and 3D `mass' maps \citep[see \eg][]{VanWaerbeke:2013eya, Chang:2015odg, Vikram:2015leg}. 
Future surveys such as the Dark Energy Survey\footnote{\url{http://www.darkenergysurvey.org/}} and LSST\footnote{\url{http://www.lsst.org/}} will be an order of magnitude wider and deeper than previous experiments. 
Therefore, accurate techniques are needed for reconstructing the convergence field (possibly via the shear \E and \B modes, as in the CMB case) in the presence of complicated sky cuts and noise over extended volumes. 

The contribution of this paper is threefold.
First, we present a new formalism to perform \E-\B reconstruction via the scale-discretised spin wavelet transform \citep{McEwen:2015s2let}.
Second, we derive a pure mode estimator for applying this technique to partial-sky data sets, so that scale- and orientation-dependent weighting can be exploited to mitigate \E-\B mixing. 
Third, we express the connection between the reconstruction formalism and other wavelet techniques applied to CMB polarisation, cosmic shear or their \E-\B modes, such as foreground cleaning.

The wavelet \E-\B reconstruction formalism presented here was used in a companion paper \citep{Rogers:2016spinsilc} to simultaneously produce improved foreground-cleaned $Q$-$U$ and \E-\B mode maps for the {\it Planck} data and simulations. The component separation method of \cite{Rogers:2016spinsilc}, Spin-SILC, is based on a internal linear combination algorithm with complex-value weights in the space of spin directional wavelets \citep{McEwen:2015s2let}; the latter allow the decomposition of the polarisation signal into \E and \B modes by separating the real and imaginary parts of the complex spin 2 wavelet coefficients. While in \cite{Rogers:2016spinsilc} Spin-SILC was applied to data and simulations on the full sky, upcoming multi-frequency CMB experiments will cover part of the sky only, at increased sensitivity; in this setting Spin-SILC, in combination with the pure estimators presented in this work, will yield a coherent pipeline allowing simultaneous treatment of  \E-\B mixing and accurate component separation.

Brief summaries of CMB polarisation, cosmic shear, and spin wavelets are given in Sec.~2. 
Pseudo and pure estimators for the recovery of \E and \B modes in harmonic and wavelet space are presented in Sec.~3. 
A demonstration on simple simulations is shown in Sec.~4, and we conclude in Sec.~5.


\section{Background}

In this section we briefly review the description of CMB polarisation and cosmic shear as spin 2 observables.
We also summarise the spin wavelet transform \citep{McEwen:2015s2let} used in the following sections to derive new wavelet space \E-\B estimators.

\subsection{Polarisation of the CMB}\label{sec:cmbpol}

The linearly-polarised light of the CMB can be described as a spin $\pm 2$ field on the sphere ${}_{\pm 2} P(\ang) = Q(\ang) \pm i U(\ang)$, with $Q$ and $U$ denoting Stokes parameters and $\ang\equiv (\theta,\phi)\in{{\rm S}^2}$ denoting angular coordinates on the sphere, with colatitude $\theta$ and longitude $\phi$ \citep[see, \eg][]{zaldarriaga:1997, Kamionkowski:1996ks}.  
In practice, the spin $\pm 2$ nature of ${}_{\pm 2}P$ implies that it is invariant under local rotations of $\pm \pi$. 
More generally, a spin $s$ field ${}_s f$ rotated locally by an angle $\chi$ satisfies ${}_s f^\prime(\ang) = e^{-is\chi}{}_s f(\ang)$, hence the invariance under $2\pi/s$ rotation. 
On the full sky, one can expand the polarisation signal using spin spherical harmonics ${}_{\pm 2} Y_{\ell m}$,
\eqn{
	{}_{\pm 2}P(\ang) = (Q \pm i U)(\ang) = \sum_{\ell m} {}_{\pm 2} a_{\ell m}\ {}_{\pm 2}Y_{\ell m}(\ang). \label{eq:qu_decomp}
}
Since $Q$ and $U$  are defined with respect to a local fixed coordinate system on the sky, one cannot define a rotation-invariant measure of the power as a function of scale, such as the power spectrum of the CMB temperature fluctuations (a spin $0$ field). 
The solution is to introduce scalar \E and pseudo scalar \B fields (referred to as \E and \B modes for brevity) defined via
\eqn{
	E_{\ell m} &= - \bigr({}_{2}a_{\ell m} + {}_{-2}a_{\ell m}\bigr) / 2 \label{eq:convEBchi1}\\
	B_{\ell m} &= \ i\bigr({}_{2}a_{\ell m} - {}_{-2}a_{\ell m}\bigr) / 2 \label{eq:convEBchi2},
}
which are spin 0 spherical harmonic coefficients defined such that 
\eqn{
  (Q\pm iU)_{\ell m} = - \bigl(E_{\ell m} \pm i B_{\ell m}\bigr).
}
These differ in parity (\E is parity-even while \B is parity-odd), fully characterise the polarisation signal, and admit rotational invariant angular power spectra, denoted by $C_\ell^{E}$ and $C_\ell^B$. 
The spatial relation between the \E-\B and \Q-\U fields is most clearly seen by introducing two other scalar and pseudo scalar fields $\epsilon$ and $\beta$ such that\footnote{Other conventions use $\chi^E$ or $\widetilde{E}$ for $\epsilon$ and $\chi^B$ or $\widetilde{B}$ for $\beta$.}
\eqn{
	 \epsilon(\ang) &=& -\frac{1}{2} \bigl[ \bar{\eth}^2(Q+iU) + \eth^2(Q-iU)  \bigr]  \\
	 &=&    - \ {\rm Re}\bigl[ \spinpm^2(Q\pm iU) \bigr]  \label{eq:epsilon_field} \\ 
 \beta(\ang) &=& \frac{i}{2} \bigl[ \bar{\eth}^2(Q+iU) - \eth^2(Q-iU)  \bigr]  \\
	 &=& \mp\ {\rm Im}\bigl[ \spinpm^2(Q\pm i U) \bigr]. \label{eq:beta_field}
}
In these expressions, ${\eth}$ and $\bar{\eth}$ are first-order differential operators know as the spin-raising and spin-lowering operators, which transform the spherical harmonics as $\eth {}_sY_{\ell m}(\ang) = \sqrt{(\ell-s)(\ell+s+1)} \ {}_{s+1}Y_{\ell m}(\ang)$ and $\bar{\eth} {}_sY_{\ell m} = -\sqrt{(\ell+s)(\ell-s+1)} \ {}_{s-1}Y_{\ell m}(\ang)$. We also make use of the compressed notation:
\eqn{
	\spinpm = 
	\begin{cases}
	 \eth & \text{if }\ +\\
	 \bar{\eth} & \text{if }\ -
	\end{cases}.
}

Importantly, $E$ and $B$ are simply rescaled versions of $\epsilon$ and $\beta$,
\eqn{
	\epsilon_{\ell m}  &=& E_{\ell m} \ N_{\ell,2} \label{eq:e_normalised} \\ 
	 \beta_{\ell m}  &=& B_{\ell m}\ N_{\ell,2}, \label{eq:b_normalised}
}
with 
\eqn{
	N_{\ell, s} = \sqrt{\frac{(\ell+s)!}{(\ell-s)!}} = \frac{1}{N_{\ell, -s}}.
}
Focusing on one doublet rather than on the other (\eg \E-\B rather that $\epsilon$-$\beta$) is only a matter of convention since their angular power spectra only differ by a factor $N^2_{\ell,2} \sim \ell^4$. 

\subsection{Cosmic shear}\label{sec:cosmicshear}

The case of cosmic shear can be obtained by a simple change of notation and the addition of a radial variable $r$ to support the inclusion of redshift or distance information.  
Specifically, the polarisation of the CMB ${}_{\pm2}P(\ang) = Q\pm iU$ involving the Stokes parameters becomes the shear ${}_{\pm2}\gamma(\ang,r) = \gamma_1 \pm i\gamma_2$.
The main underlying scalar fields of interest are no longer \E and \B (or $\epsilon$ and $\beta$) but rather the lensing potentials ${\phi}^E$ and ${\phi}^B$ explicitly defined below.
While the shear induced by gravitational lensing produces an \E-mode signal only, the \B mode is a powerful check for data systematics and might also be created in the context of non-standard cosmological models; therefore we include both ${\phi}^E$ and ${\phi}^B$ in the formalism below.
More details about 3D cosmic shear and its formulation in terms of spin observables can be found in \cite{heavens:2003, castro:2005, Heavens:2006uk, Leistedt:2015vwa}.

When a signal on the sphere is extended with a radial dimension $r$, the natural harmonic transform is the Fourier-Bessel transform, obtained by complementing the spin spherical harmonics ${}_sY_{\ell m}$ with spherical Bessel functions $j_\ell(kr)$. 
A 3D field ${}_sf(\ang,r)$ with (angular) spin $s$ symmetries is transformed as
\eqn{
	{}_sf_{\ell m}(k) &=&  \int_{{\rm S}^2}  {\rm d}\Omega(\ang) \sqrt{\frac{2}{\pi}}  \int_{\mathbb{R}^+}   {\rm d} r r^2  {}_sf(\ang,r)  \ {}_sY^*_{\ell m}(\ang)  j^*_\ell(kr)\label{fourierbesselanalysis} \\
	{}_sf(\ang,r) &=&  \sum_{\ell m} \sqrt{\frac{2}{\pi}} \int_{\mathbb{R}^+} {\rm d} k k^2 \ {}_sf_{\ell m}(k)  \ {}_sY_{\ell m}(\ang)   j_\ell(kr) \label{fourierbesselsynthesis} ,
}
with $_sf_{\ell m}(k)$ the Fourier-Bessel coefficients. 
This notation is used in the remainder of this section.

Gravitational lensing generates distortions in the observations of a background field \citep[for a review of the physics of gravitational lensing, see \eg][]{Bartelmann:1999yn}. 
In the weak lensing regime (\ie away from the critical curve of lensing masses, where there are no multiple images of sources) three types of distortion can be produced: the size magnification; the shear; and the flexion.
Here we focus on the shear, usually decomposed into real and imaginary parts, and related to the lensing potential via
\eqn{
	{}_{\pm2} \gamma(\ang, r) \ = \ \gamma_1(\ang,r) \pm i\gamma_2(\ang,r) \ =\ \spinpm^2\bigl(\phi^E(\ang,r) \pm i \phi^B(\ang,r)\bigr)/2.
}
Note that these are distinct from the standard \E and \B fields (due to a normalisation), which are defined in terms of Fourier-Bessel coefficients by
\eqn{
	{}_{\pm2}\gamma_{\ell m}(k) \ =\ - \bigl(E_{\ell m}(k) \pm {\rm i} B_{\ell m}(k)\bigr).
}
Thus,
\eqn{
	E_{\ell m}(k) &=& -\bigl( {}_{2}\gamma_{\ell m}(k) + {}_{- 2}\gamma_{\ell m}(k)\bigr) /2\\
	B_{\ell m}(k) &=& i\bigl( {}_{2}\gamma_{\ell m}(k) - {}_{- 2}\gamma_{\ell m}(k) \bigr) /2 .
}
The only difference with CMB polarisation described in the previous section is the extra radial Fourier mode $k$. 
However, a different convention is typically used for the equivalent of the $\epsilon$ and $\beta$ fields, denoted by
\eqn{
	{\varphi}^E(\ang,r)  &=& \bigl( \bar{\eth}^2 {}_2 \gamma +  {\eth}^2 {}_{-2} \gamma \bigr)\quad \\
	{\varphi}^B(\ang,r)  &=& -{\rm i} \bigl( \bar{\eth}^2 {}_2 \gamma - {\eth}^2 {}_{-2} \gamma \bigr).
}
Thus, these are connected to the lensing potential \E and \B fields via
\eqn{
	{\varphi}^E_{\ell m}(k) &=& (N_{\ell,2})^2 {\phi}^E_{\ell m}(k) \\
	{\varphi}^B_{\ell m}(k) &=& (N_{\ell,2})^2 {\phi}^B_{\ell m}(k),
}
and to the \E and \B-mode fields via
\eqn{
	\phi^E_{\ell m}(k) &=& - 2 N_{\ell, -2} E_{\ell m}(k)\\
	\phi^B_{\ell m}(k) &=& - 2 N_{\ell, -2} B_{\ell m}(k).
}

Similarly to the CMB polarisation case, the main observable measured in data is ${}_{\pm2} \gamma(\ang, r)$, and one is interested in mapping the \E and \B-mode fields (or equivalently ${\varphi}^E$ and ${\varphi}^B$) to constrain the lensing potential(s) $\phi^E$ ($\phi^B$) predicted by cosmological models.

\subsection{Spin scale discretised directional wavelets}\label{sec:waveletsummary}

Spin, directional, scale-discretised wavelets on the sphere that support exact reconstruction have been constructed in \citet{McEwen:2015s2let} (introduced briefly in \citealt{2014IAUS..306...64M,2015arXiv150203120L}) and extend to spin functions the scalar wavelets detailed in \cite{Wiaux:2007ri, Leistedt:2012gk, McEwen:2013tpa}. Scale-discretised wavelets satisfy excellent localisation properties, both in the spatial and harmonic domains \citep{mcewen:2015s2let_localisation}.
In other words, corresponding wavelet coefficients extract signal content localised in space, scale or frequency, and direction, which makes wavelets a powerful analysis tool.
The wavelet transform of a spin $s$ function ${}_sf$ on the sphere is defined as 
\eqn{
	W_{{}_sf}^{{}_s\Psi^j}(\rho) &=& \int_{{\rm S}^2} {\rm d}\Omega(\ang)\  {}_sf(\ang) \ [ \mathcal{R}_\rho\ {}_s\Psi^j ]^*(\ang) \quad j=J_0,\ldots,J \label{eq:wavanalysis} \\
	W_{{}_sf}^{{}_s\Phi}(\ang') &=& \int_{{\rm S}^2} {\rm d}\Omega(\ang)\  {}_sf(\ang) \ [ \mathcal{R}_{\ang'}\ {}_s\Phi ]^*(\ang). \label{eq:wavanalysis_scal}
}
In these equations, $\rho\equiv(\alpha,\beta,\gamma)\in {\rm SO}(3)$ specifies a 3D rotation, characterised by the rotation group ${\rm SO}(3)$ and parameterised by the Euler angles $(\alpha,\beta,\gamma)$. 
Furthermore, ${\rm d}\Omega(\ang)=\sin\theta\,{\rm d}\phi\,{\rm d}\theta$ and ${\rm d}\mu(\rho)=\sin\beta\,{\rm d}\beta\,{\rm d}\alpha\, {\rm d}\gamma$ are the usual measures on the sphere ${\rm S}^2$ and on the rotation group ${\rm SO}(3)$. 
Finally, $\mathcal{R}_\rho$ and $\mathcal{R}_\ang$ are rotation operators on ${\rm SO}(3)$ and ${\rm S}^2$, respectively.
These quantities are defined in further detail in \cite{McEwen:2015s2let}. 

Wavelet coefficients $W_{{}_sf}^{{}_s\Psi^j}(\rho)$ are defined as a function of scale $j=J_0,\ldots,J$  as the directional convolution of ${}_sf$ with the wavelet ${}_s\Psi^j$ ($J_0$ and $J$ are defined precisely below to ensure an invertible transform). 
 These typically do not capture the large scale information content of the signal \citep{Wiaux:2007ri, Leistedt:2012gk, McEwen:2013tpa, McEwen:2015s2let}, which justifies the introduction of scaling function coefficients $W_{{}_sf}^{{}_s\Phi}(\ang')$ of \equref{eq:wavanalysis_scal}, defined as the axisymmetric convolution of ${}_sf$ with the scaling function ${}_s\Phi$.

The input signal can be reconstructed from its wavelet coefficients by
\eqn{
	{}_sf(\ang) &=& \int_{{\rm S}^2} {\rm d}\Omega(\ang') \ W_{{}_sf}^{{}_s\Phi}(\ang') \ [ \mathcal{R}_{\ang'}\ {}_s\Phi ](\ang)  \label{eq:wavsynthesis}\\
	&& + \ \sum_{j=J_0}^J \int_{\rm SO(3)} {\rm d}\mu(\rho) \ W_{{}_sf}^{{}_s\Psi^j}(\rho) \ [ \mathcal{R}_\rho\ {}_s\Psi^j ](\ang)\nonumber.
}

In order for this transform to be invertible, \ie for the coefficients to capture all the information content of ${}_sf$, the wavelets ${}_s\Psi^j$ and the scaling function ${}_s\Phi$ must be chosen such that their spherical harmonic coefficients satisfy  
\eqn{
	\frac{4\pi}{2\ell+1} | {}_s\Phi_{\ell 0}|^2 \ + \ \frac{8\pi^2}{2\ell+1} \sum_{j=J_0}^J \sum_{m=-\ell}^\ell| {}_s\Psi_{\ell m}^j |^2 \ = \ 1, \quad \forall \ell.
}
As in \cite{McEwen:2015s2let}, we follow the construction of scale discretised wavelets \citep{wiaux:2007:sdw, mcewen:2015s2let_localisation} and define wavelets in harmonic space by
\eqn{
	{}_s\Psi_{\ell m}^j = \sqrt{\frac{2\ell+1}{8 \pi^2}} \kappa^j_\ell \: \zeta_{\ell m}. \label{eq:wavharmndef}
}
In this construction, $\kappa^j_{\ell}$ controls the scales probed by the $j$th wavelet: it peaks at $\ell = \lambda^j$ and is compact (\eg non zero) for $\lambda^{j-1} \leq \ell \leq \lambda^{j+1}$ only.
Similarly, $\zeta_{\ell m}$ controls the directionality of the wavelet and is constructed such that $\sum_m |\zeta_{\ell m}|^2=1$.
For more details about the computation of these coefficients see \cite{wiaux:2007:sdw, leistedt:s2let_axisym, mcewen:2015s2let_localisation, McEwen:2015s2let}. 
Note that the wavelets can probe directional structure and hence wavelet coefficients live on the rotation group, while the scaling function is axisymmetric and hence scaling coefficients live on the sphere.

Assuming that the signal ${}_{s}f$ is band-limited at $L$ (\ie \mbox{${}_{s}f_{\ell m} = 0,\ \forall \ell \ge L$}), the maximum scale $J$ is set to \mbox{$J = \lceil \log_\lambda(L-1) \rceil$} to capture all the information and achieve exact reconstruction. 
The minimal scale $J_0$ can be chosen arbitrarily, provided the wavelet transform is complemented with a scaling function capturing the large scale, low-frequency content of the signal (formally, to analyse ${}_{s}f_{\ell m}$ with $\ell \leq \lambda^{J_0}$). 
As a side note, spin $s$ signals have ${}_{s}f_{\ell m} = 0, \ \forall \ell < s$, which may alleviate the need for a scaling function depending on the chosen parameters (\eg $\lambda=2, J_0=0$).

Fast algorithms to compute scale-discretised wavelet transforms were presented in \cite{McEwen:2015s2let, 2013SPIE.8858E..0IM, mcewen:2006:fcswt, wiaux:2005c}. We adopt the latest developments presented in \cite{McEwen:2015s2let}, which rely on the fast spherical and Wigner transforms of \cite{mcewen:fssht} and \cite{2015ISPL...22.2425M}, respectively.  
Furthermore, the wavelet transform is made theoretically exact in practice by making use of the sampling theorems of \cite{mcewen:fssht} and \cite{2015ISPL...22.2425M} on the sphere and the rotation group, respectively.
These allow one to evaluate the integrals of the forward and inverse transforms of \equrefss{eq:wavanalysis}{eq:wavanalysis_scal}{eq:wavsynthesis} with exact quadrature rules. 
In other words, no approximation is made in evaluating these integrals, and the only numerical errors are due to manipulating floating point numbers, which can only be represented at finite precision in practice (up to 16 decimals for double precision floats).

The wavelet transform depends on two main parameters: $\lambda$, controlling the size of the $\kappa^j_\ell$ windows and therefore setting the scales (harmonic multipoles $\ell$) probed by each wavelet; and $N$ the azimuthal band-limit, setting $\zeta_{\ell m}$.
This parameter in fact controls the number of directions, which correspond to a discretisation of the azimuthal rotation angle $\gamma$ into $\gamma_n$ with $n=1,\ldots,N$ indexing the directions.\footnote{While this implies that only $N$ directions are stored, by steerability of the wavelets all other directions are in fact accessible by the transform and can be reconstructed from the $N$ stored  (see \citealt{McEwen:2015s2let} for more details about steerability).}

\section{Pure \E-\B separation with spin wavelets}

In this section we introduce a formalism to perform \E-\B reconstruction with spin wavelets.
We describe the (standard) harmonic space and the (new) wavelet space \E-\B reconstruction methods, both in the case of full sky and partial sky coverage.
We defer the discussion of the explicit construction of masks for these estimators to \secref{sec:results}.
We adopt the CMB notation presented in \secref{sec:cmbpol} but all results can be adapted to cosmic shear by performing the notational changes described in \secref{sec:cosmicshear}.
Note that in the following derivations we present results for wavelet coefficients only and not also scaling coefficients (since the latter is identical up to a notational change).

\subsection{Recovering \E-\B modes on the full sky}

In the full sky setting, \E and \B modes can be computed from $Q\pm iU$ simply via \equrefs{eq:convEBchi1}{eq:convEBchi2}, noting \equref{eq:qu_decomp}.  We present an analogous connection in the wavelet formalism, introduced briefly in \cite{McEwen:2015s2let, 2014IAUS..306...64M, 2015arXiv150203120L}.
We start by noting the spatial connection between $Q\pm iU$ and \E-\B of \equrefs{eq:epsilon_field}{eq:beta_field}, expanding $Q\pm iU$ into its spin wavelet representation of \equref{eq:wavsynthesis}, propagating the spin operators and then noting the forward wavelet transform of \equref{eq:wavanalysis}. Some simple algebra then allows us to connect the spin $0$ wavelet transform of $\epsilon/\beta$ (with wavelets ${}_0\Psi^j$) to the spin $\pm2$ wavelet transform of $Q\pm iU$ (using wavelets ${}_{\pm2}\Psi^j$), and to write
\eqn{
	W_{\epsilon}^{{}_0\Psi^j} (\rho) \hspace*{-1mm} &=& \hspace*{-1mm} \int_{{\rm S}^2}{\rm d}\Omega(\ang) \ \epsilon(\ang) \ {}_0\wav_\rho{}^\ast(\ang) \ = \   - {\rm Re} \left[  W_{Q\pm iU}^{{}_{\pm2}\Psi^j} (\rho) \right] \label{equ:wavEBsepfs1} \\
	W_{\beta}^{{}_0\Psi^j} (\rho) \hspace*{-1mm} &=&  \hspace*{-1mm} \int_{{\rm S}^2}{\rm d}\Omega(\ang) \ \beta(\ang) \ {}_0\wav_\rho{}^\ast(\ang)   \ = \ \mp {\rm Im} \left[  W_{Q\pm iU}^{{}_{\pm2}\Psi^j} (\rho) \right]. \label{equ:wavEBsepfs2}
}
In other words, \E and \B modes can be recovered by computing a \emph{spin} wavelet transform of ${}_{\pm 2} P = Q \pm i U$, followed by \emph{scalar} inverse wavelet transforms of the real and imaginary parts of the spin wavelet coefficients.  Equivalently, we have
\equ{
	W_{{}_{\pm 2} P}^{{}_{\pm2}\wav}(\rho) 
	= \int_{{\rm S}^2}{\rm d}\Omega(\ang) \ {}_{\pm 2} P(\ang)\  {}_{\pm2}\wav_\rho{}^\ast(\ang) 
	= \Bigr[- W_{\epsilon}^{{}_0\wav} \mp i W_{\beta}^{{}_0\wav} \Bigl](\rho) .
}

Importantly, the connections outlined above are only satisfied if ${}_0\Psi^j$ and ${}_{\pm2}\Psi^j$ are constructed such that
\eqn{
	{}_0\Psi^j (\ang)  = \eth_\mp^2[ {}_{\pm2}\Psi^j ](\ang),
}
or equivalently 
\eqn{
	{}_0\Psi^j_{\ell m} = N_{\ell, 2}  \ {}_{\pm2}\Psi^j_{\ell m}  ,
	}
and the scalar wavelets ${}_0\Psi^j$ are real.  In other words, the scalar wavelet must be a spin lowered/raised version of the spin wavelet used to analyse ${}_{\pm 2} P$.
As a consequence, one can define ${}_0\Psi^j_{\ell m}$ or ${}_{2}\Psi^j_{\ell m}$ by \equref{eq:wavharmndef}, and rescale the other with $N_{\ell, 2}$. 
In our implementation, we set ${}_{2}\Psi^j_{\ell m}=\sqrt{\frac{2\ell+1}{8 \pi^2}} \kappa^j_\ell \zeta_{\ell m}$.

In summary, the spin wavelet transform of the polarisation field $Q\pm iU$ is connected to the scalar wavelet transform of the \E and \B fields using spin raised or lowered wavelets.  This property is a specific consequence of the spin scale-discretised wavelet construction, where the wavelet coefficients of a spin function are themselves scalar functions \citep{McEwen:2015s2let}, and is not necessarily satisfied in alternative spin wavelet constructions.
This connection provides a natural framework for reconstructing the \E-\B modes from $Q\pm iU$, allowing one to exploit the spatial, directional and harmonic localisation of the wavelets \citep{mcewen:2015s2let_localisation}.
For example, this can be useful for efficiently mitigating \E-\B mixing using scale- and direction-dependent weighting and masking, and for interfacing with other wavelet space algorithms as discussed in \secref{sec:discussion}

We now examine the connection between the Stokes parameters \Q and \U and the \E-\B modes in the case of incomplete sky coverage.

\subsection{\E-\B separation on the cut sky: leakage and pure modes}

We now consider the setting where the $Q \pm iU$ signal is restricted to a portion of the sky, defined via a scalar mask $M(\ang)$, with real values in the range $[0,1]$. 
When dealing with partial sky coverage, the previous relations (including \equrefs{eq:convEBchi1}{eq:convEBchi2} and \equrefs{equ:wavEBsepfs1}{equ:wavEBsepfs2}) are no longer accurate in practice due to finite resolution, \ie they do not yield unbiased estimates of the \E and \B maps or their spherical harmonic coefficients. 
More precisely, there is {\it leakage} or {\it mixing} from \E to \B, and conversely. 
This is a well-known problem arising from the fact that polarisation can only be uniquely decomposed into \E and \B modes on a manifold without boundary. 
In fact, if we defined \E  and \B  mode fields as having vanishing $\beta$ and $\epsilon$ components respectively (which is what we have implicitly assumed so far), cut-sky polarisation signals admit a set of {\it ambiguous} modes that satisfy the definitions of \E or \B (curl-free or divergence-free). 
The solution is to estimate $\epsilon$ and $\beta$ modes that only contain pure \E and \B modes respectively \cite[\eg][]{Smith2006pureEB, SmithZaldarriaga2007pureEB, Grain2012crosspure, Ferte2014pureEB}. Recall, pure \E (\B) modes are orthogonal to all \B (\E) modes in the sky region of interest. 
Pure \B mode estimators are of considerable interest since the CMB \E modes are orders of magnitude larger than \B, and any \E to-\B leakage significantly affects the \B mode estimates. 

In what follows, we present standard (pseudo) and pure mode estimators in both harmonic and wavelet space. We adopt the following generic notation for the mask and its derivatives,
\eqn{
	{}_0 M = M, \quad {}_{\pm 1} M = \spinpm M, \quad {}_{\pm 2} M = \spinpm^2 M,
}
as well as their products with the Stokes parameters,
\eqn{
	{}_{\pm 2}\widetilde{P} = {}_0 M {}_{\pm 2}P, \quad {}_{\pm 1}\widetilde{P} = {}_{\mp 1} M {}_{\pm 2}P, \quad {}_{\pm 0}\widetilde{P} = {}_{\mp 2} M {}_{\pm 2}P.
}
Finally, we introduce the quantities
\eqn{
	{}_s\mathcal{E}_{\ell, m} &=& -\frac{1}{2} \Bigr[ {}_{s}\widetilde{P}_{\ell m} + {}_{-s}\widetilde{P}_{\ell m} \Bigl] \label{eq:epsilon_cal_lm} \\
	{}_s\mathcal{B}_{\ell, m} &=& \ \ \frac{i}{2} \Bigr[ {}_{s}\widetilde{P}_{\ell m} + {}_{-s}\widetilde{P}_{\ell m} \Bigl] \label{eq:beta_cal_lm}.
}
Throughout we follow the convention that quantities adorned with a tilde (\ie $\widetilde{\cdot}\:$) are masked or pseudo quantities, while those adorned with a hat (\ie $\widehat{\cdot}\:$) are pure quantities.
Note that for the wavelet \E-\B reconstruction the smoothing of the mask can depend on scale and direction (\ie scale- and direction-dependent weighting and masking can be exploited), in which case the quantities defined above require extra superscripts (as introduced later when needed). 

\subsubsection{Harmonic space pseudo estimator}

In the standard harmonic space pseudo \E-\B estimator one analyses the spherical harmonic coefficients of the masked polarisation signal, ${}_{\pm 2}\widetilde{P}$. Using the notation above, their spherical harmonic coefficients read \citep[\eg][]{kim:2011}
\eqn{
	\widetilde{E}^{\rm harm}_{\ell m} &=& {}_2\mathcal{E}_{\ell, m} \label{eq:e_harmonic_pseudo}\\
	\widetilde{B}^{\rm harm}_{\ell m} &=& {}_2\mathcal{B}_{\ell, m} \label{eq:b_harmonic_pseudo}.
}
Their power spectra relate to the power spectra of the true, full sky \E and \B modes through the widespread pseudo power spectrum estimator \citep[\eg][]{hivon:2002}
\eqn{
	\left( \begin{array}{l} \widetilde{C}^E_\ell \\  \widetilde{C}^B_\ell \end{array} \right) 
	= \sum_{\ell'} 
	\left( \begin{array}{cc} \widetilde{M}^{+}_{\ell \ell'} & \widetilde{M}^{-}_{\ell \ell'} \\ \widetilde{M}^{-}_{\ell \ell'} & \widetilde{M}^{+}_{\ell \ell'} \end{array} \right)
	\left( \begin{array}{l} {C}^E_\ell \\  {C}^B_\ell \end{array} \right),
} 
with the coupling matrices
\eqn{
	\widetilde{M}^{\pm}_{\ell \ell'} \ = \ \frac{2\ell'+1}{16\pi} \sum_{\ell''} (2\ell''+1) W_{\ell''} \Bigl[ J_0^{\pm}(\ell,\ell',\ell'') \Bigr]^2 ,\label{eq:couplingpseudo}
}
where
\eqn{
	J_s^{\pm}(\ell,\ell',\ell'') \ =\ \left( \begin{array}{ccc} \ell & \ell' & \ell'' \\ s-2 & 2 & -s \end{array} \right) \ \pm\ \left( \begin{array}{ccc} \ell & \ell' & \ell'' \\ 2-s & -2 & s \end{array} \right) 
}
and $W_{\ell''}$ is the angular power spectrum of the mask ${}_0M(\ang)$. 

Inverting this system yields unbiased estimates of the full sky power spectra of interest ${C}^E_\ell$ and ${C}^B_\ell$. 
However, in the case of the CMB the variance of the \B mode power spectrum estimates is prohibitively large due to the leakage from the \E modes which have a much larger amplitude than the \B modes.
This can be seen from the off-diagonal elements $\widetilde{M}^{-}$ which capture the mixing between \E and \B power spectrum due to the ambiguous modes.

The pure estimator, which aims to resolve this issue, consists of writing an estimator for the masked \E and \B fields directly (or equivalently estimators of the masked fields $\epsilon$ and $\beta$).

\subsubsection{Harmonic space pure estimator}

Assuming that the mask and its derivatives vanish at its boundaries, one can show that the construction
\eqn{
	\widehat{E}^{\rm harm}_{\ell m} &=&  {}_2\mathcal{E}_{\ell, m}  \ +\ 2\ N_{\ell, -2}\ N_{\ell, 1}\ {}_1\mathcal{E}_{\ell, m} \ +\ N_{\ell, -2}\ {}_0\mathcal{E}_{\ell, m} \label{eq:e_harmonic_pure}\\
	\widehat{B}^{\rm harm}_{\ell m} &=&  {}_2\mathcal{B}_{\ell, m}  \ +\ 2\ N_{\ell, -2}\ N_{\ell, 1}\ {}_1\mathcal{B}_{\ell, m} \ +\ N_{\ell, -2}\ {}_0\mathcal{B}_{\ell, m} \label{eq:b_harmonic_pure}
}
is an extension of the previous pseudo spectrum approach that includes extra terms to cancel out ambiguous \E-\B modes and thus the \E-\B leakage. 
In this case, a valid power spectrum estimator is identical to the pseudo spectrum case but with
\eqn{
	\widehat{M}^{\pm}_{\ell \ell'} & = & \frac{2\ell'+1}{16\pi} \sum_{\ell''} (2\ell''+1) W_{\ell''}  \Bigl[ \	N_{\ell, -2}\ N_{\ell'', 2} \ J_2^{\pm}(\ell,\ell',\ell'') \nonumber\\
	&+& \hspace*{-1mm} 2\ N_{\ell, 1} \ N_{\ell, -2}\ N_{\ell'', 1} \ J_1^{\pm}(\ell,\ell',\ell'') \ +\ J_0^{\pm}(\ell,\ell',\ell'') \	\Bigr]^2.\label{eq:couplingpure}
}
As before, these expressions are only valid if the mask (or an apodised, weighted version of it) satisfies the Dirichlet and Neumann boundary conditions (\ie that the mask and its derivative vanish at the boundaries of the mask).
In fact, the construction of this mask is critical: it determines which ambiguous modes are cancelled out.
Consequently, even though pure estimators yield unbiased \E and \B maps and power spectra, their variance and information content critically depends on the mask.
For this reason, the construction of optimal masks which minimise both the loss of information due to mode removal and the leakage-induced variance is of central interest.
However, the variance of pure estimators is smaller than that of standard pseudo estimators by construction, and mask optimisation has been extensively studied previously in the context of power spectra estimation \citep[\eg][]{Smith2006pureEB, SmithZaldarriaga2007pureEB, Grain2012crosspure, Ferte2014pureEB}. 

\subsubsection{Wavelet space pseudo estimator}

We now examine how the wavelet space formalism is modified in the cut-sky setting. 
If no corrections are applied and the masked data are analysed as in the full sky setting, the result is analogous to the standard harmonic space pseudo estimator and reads
\equ{
	{W}_{{}_{\pm 2} \widetilde{P}}^{{}_2\wav}(\rho) 
	= \int_{{\rm S}^2}{\rm d}\Omega(\ang) \ {}_{\pm 2} \widetilde{P}(\ang)\  {}_2\wav_\rho{}^\ast(\ang) ,
}
where pseudo wavelet estimators of \E and \B may be recovered via
\eqn{
  \widetilde{W}_{\epsilon}^{{}_0\Psi^j} (\rho) \hspace*{-1mm} &=&  - {\rm Re} \left[  {W}_{{}_{\pm 2} \widetilde{P}}^{{}_{\pm2}\Psi^j} (\rho) \right] \\
  \widetilde{W}_{\beta}^{{}_0\Psi^j} (\rho) \hspace*{-1mm} &=&  \mp {\rm Im} \left[  {W}_{{}_{\pm 2} \widetilde{P}}^{{}_{\pm2}\Psi^j} (\rho) \right],
}
from which maps can be computed by inverse scalar wavelet transforms.

Similar conclusions to the harmonic setting can be reached regarding spectral estimators: a pseudo spectrum estimator can be derived but leads to a large variance in the \B modes due to significant \E-to-\B leakage.

For notational simplicity, we have assumed here that the same mask $M(\ang)$ is applied for all wavelet scales $j$.  However, one of the key advantages of wavelet \E-\B separation is the ability to apply scale- and orientation-dependent masking, \ie masks that vary with scale and orientation.  We present this generalisation  in \secref{sec:multiscalemask}.

\subsubsection{Wavelet space pure estimator}

As in the harmonic case, the  wavelet pseudo estimator can be extended to cancel out the ambiguous modes and yield higher quality \E-\B separation.
The starting point is to write the wavelet coefficients of the masked $\epsilon$ and $\beta$ fields by
\eqn{
	&&\hspace*{-7mm}  \widehat{W}_{\epsilon}^{{}_0\Psi}(\rho) = \int_{{\rm S}^2} {\rm d}\Omega(\ang) \ \epsilon(\ang) \ M(\ang) \ [ \mathcal{R}_\rho\ {}_0\Psi^j ]^*(\ang) \\
	&&	 \hspace*{5mm} = - {\rm Re} \int_{{\rm S}^2} {\rm d}\Omega(\ang) \  {\eth}^2_\mp(Q\pm iU)(\ang) \ M(\ang) \ [ \mathcal{R}_\rho\ {}_0\Psi^j ]^*(\ang)  \\
	&&\hspace*{-7mm}  \widehat{W}_{\beta}^{{}_0\Psi^j}(\rho) = \int_{{\rm S}^2} {\rm d}\Omega(\ang) \ \beta(\ang) \ M(\ang) \ [ \mathcal{R}_\rho\ {}_0\Psi^j ]^*(\ang) \\
	&&	 \hspace*{5mm} = \mp\ {\rm Im} \int_{{\rm S}^2} {\rm d}\Omega(\ang) \  {\eth}^2_\mp(Q\pm iU)(\ang) \ M(\ang) \ [ \mathcal{R}_\rho\ {}_0\Psi^j ]^*(\ang)  ,
}
which by definition are pure modes.  Again, for notational simplicity, we consider a single mask $M(\ang)$; we generalise to a scale- and orientation-dependent mask in \secref{sec:multiscalemask}.

To express these pure estimators in terms of the masked versions of the observable Stokes parameters $Q \pm i U$, we move the action of $\eth^2_\mp$ from $Q\pm iU$ to the masked basis function, here $M \ ( \mathcal{R}_\rho\ {}_0\Psi^j )$. 
This can be performed by integration by parts, realising $\eth_\pm$ are covariant differential operators on the sphere, so the Leipzig rules on derivatives (as well as integration by parts) apply. 
By assuming the Dirichlet and Neumann boundary conditions, \ie that the mask and its derivative vanish at the boundaries of the mask $M(\ang)$, one can obtain
\eqn{
	W_ {\epsilon}^{{}_0\Psi^j}(\rho) = - {\rm Re} \int_{{\rm S}^2} {\rm d}\Omega(\ang) \ (Q\pm iU)(\ang) \ {\eth}^2_\mp[ M(\ang) (\mathcal{R}_\rho\ {}_0\Psi^j)(\ang) ] \ \ \\
	W_ {\beta}^{{}_0\Psi^j}(\rho) = \mp\ {\rm Im} \int_{{\rm S}^2} {\rm d}\Omega(\ang) \ (Q\pm iU)(\ang) \ {\eth}^2_\mp [ M(\ang) (\mathcal{R}_\rho\ {}_0\Psi^j)(\ang) ] .
}
The last quantities in the above expressions can be expanded as
\eqn{
	&& \hspace{-8mm} \spinpm^2[ M\ (\mathcal{R}_\rho\ {}_0\Psi^j) ] \nonumber \\
   &=& M\ \spinpm^2(\mathcal{R}_\rho\ {}_0\Psi^j)
	+  2 \ \spinpm M\ \spinpm(\mathcal{R}_\rho\  {}_0\Psi^j)   
	+  \spinpm^2 M\ (\mathcal{R}_\rho\ {}_0\Psi^j)  \\
	 &=&   {}_0M  \ (\mathcal{R}_\rho\ {}_{\pm2}\Upsilon^j)
	\ +\   2 \ {}_{\pm1}M \ (\mathcal{R}_\rho\ {}_{\pm1}\Upsilon^j)
	\ + \  {}_{\pm2}M\  (\mathcal{R}_\rho\ {}_0\Upsilon^j),
}
where the angular dependency $\ang$ was omitted for concision. 
In this expression we define the spin adjusted wavelets by \mbox{${}_{\pm s}\Upsilon^j = \spinpm^s({}_0\Psi^j)$}.  Note that these wavelets differ from the original wavelets ${}_{\pm 2}\Psi$ due to differing normalisations (\eg ${}_{2}\Upsilon^j = \eth^2 \bar{\eth}^2 {}_{2}\Psi$).  Consequently, the spherical harmonic coefficients of the spin adjusted wavelets are given by
\eqn{
	{}_0\Psi^j_{\ell m}  = \frac{ {}_{\pm2}\Upsilon^j_{\ell m}}{ N_{\ell, 2}} = \frac{ {}_{\pm1}\Upsilon^j_{\ell m}}{ (\pm1) N_{\ell, 1}} = {}_0\Upsilon^j_{\ell m}.
}
The pure \E-\B estimators in wavelet space then follow and read
\eqn{
	\widehat{W}_{\epsilon}^{{}_0\Psi^j}(\rho) &=&  -\ {\rm Re} \left[  W_{{}_{\pm2}\widetilde{P}}^{{}_{\pm2}\Upsilon^j}(\rho) + 2 W _{{}_{\pm1}\widetilde{P}}^{{}_{\pm1}\Upsilon^j}(\rho) + W_{{}_0\widetilde{P}}^{{}_0\Upsilon^j}(\rho) \right] \label{eq:waveletEpure}\\
	\widehat{W}_{\beta}^{{}_0\Psi^j}(\rho) &=&  \mp\ {\rm Im} \left[  W_{{}_{\pm2}\widetilde{P}}^{{}_{\pm2}\Upsilon^j}(\rho) + 2 W_{{}_{\pm1}\widetilde{P}}^{{}_{\pm1}\Upsilon^j}(\rho) + W_{{}_0\widetilde{P}}^{{}_0\Upsilon^j}(\rho) \right] \label{eq:waveletBpure},
}
from which maps can be computed by inverse scalar wavelet transforms.
Here we set ${}_{0}\Psi^j_{\ell m}=\sqrt{\frac{2\ell+1}{8 \pi^2}}\kappa^j_\ell \zeta_{\ell m}$.

The wavelet pure \E-\B estimators are analogous to the harmonic space pure estimators \cite[\eg][]{Lewis:2003an, bunn:2003, Smith2006pureEB, SmithZaldarriaga2007pureEB, Grain2012crosspure, Ferte2014pureEB}, which is not surprising; however there exist some subtle differences. The first term on the right hand side of \equrefs{eq:waveletEpure}{eq:waveletBpure} is the result of a spin $\pm 2$ wavelet transform of the masked $Q\pm iU$ signal.  Notice that this term is not identical to the wavelet pseudo estimator since a different (renormalised) wavelet is used.  Up to a normalisation, however, the first term captures the pseudo estimator contribution to the pure mode, while the second and third terms cancel out ambiguous modes. These second and third terms are spin $1$ and spin $0$ wavelet transforms of $Q\pm iU$ masked with the first and second derivatives of the mask, respectively.  

Consequently, to construct wavelet pure \E and \B mode estimators it is necessary to perform not only spin 2 and scalar (spin 0) wavelet transforms but also spin 1 wavelet transforms.  Fortunately, the construction of \cite{McEwen:2015s2let} yields a spin wavelet formalism for arbitrary spin, which is not the case for any other spin wavelet construction on the sphere. Moreover, changing the spin number does not alter the computation time of the fast algorithm presented in \cite{McEwen:2015s2let} to compute spin wavelet transforms.

\subsubsection{Scale- and orientation-dependent masking}
\label{sec:multiscalemask}

\begin{figure*}
\includegraphics[width=5.8cm]{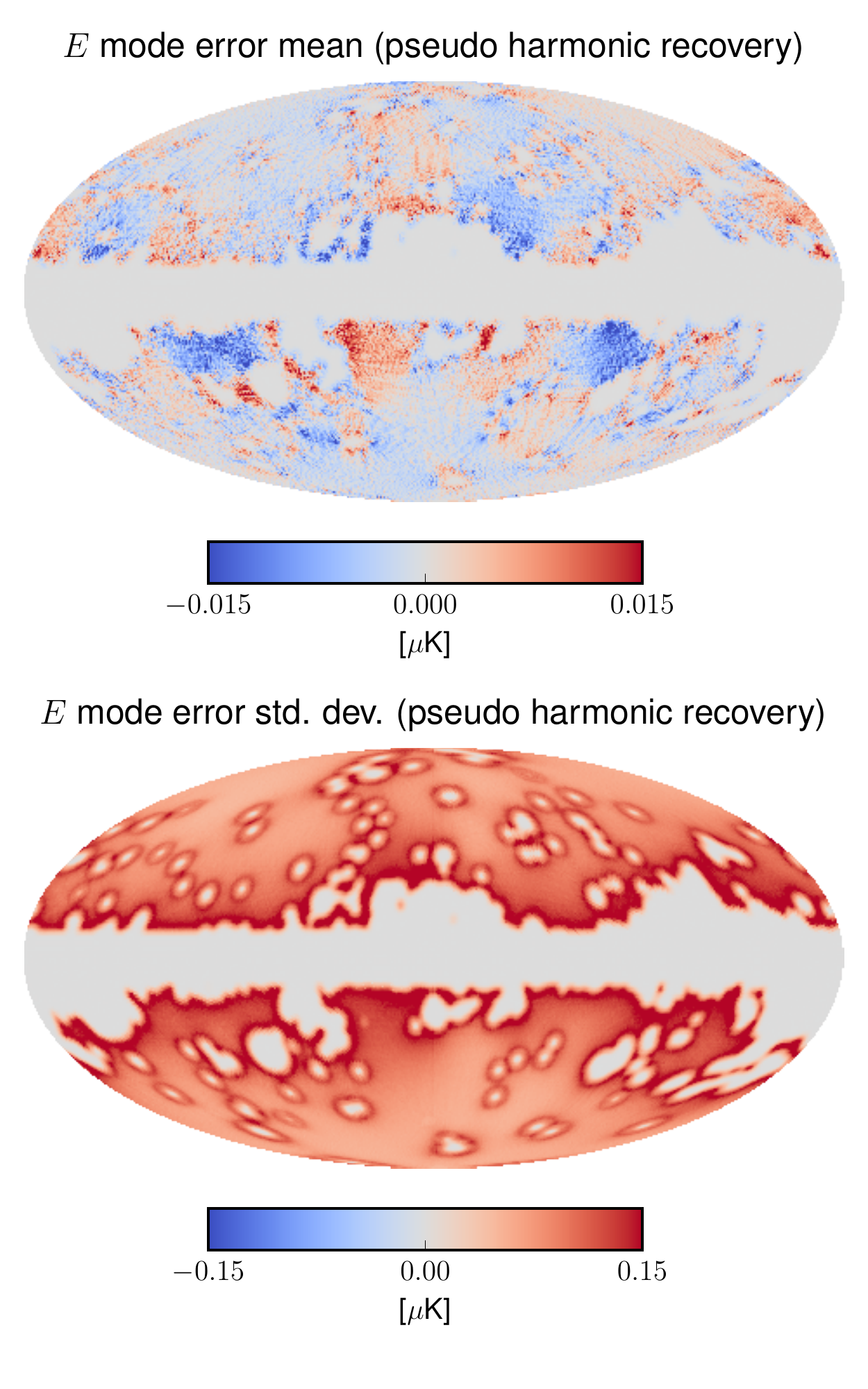}
\includegraphics[width=5.8cm]{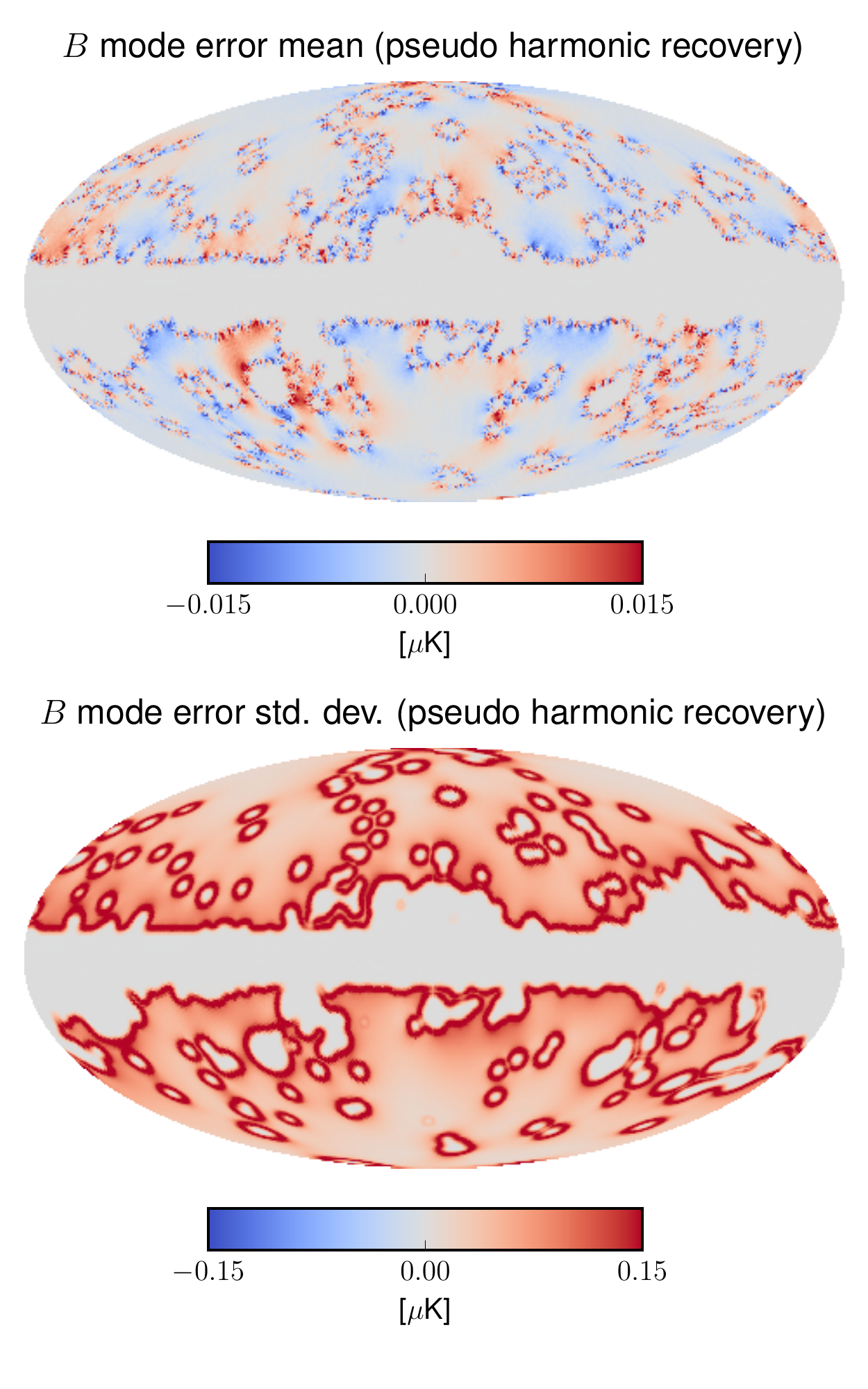}
\includegraphics[width=5.8cm]{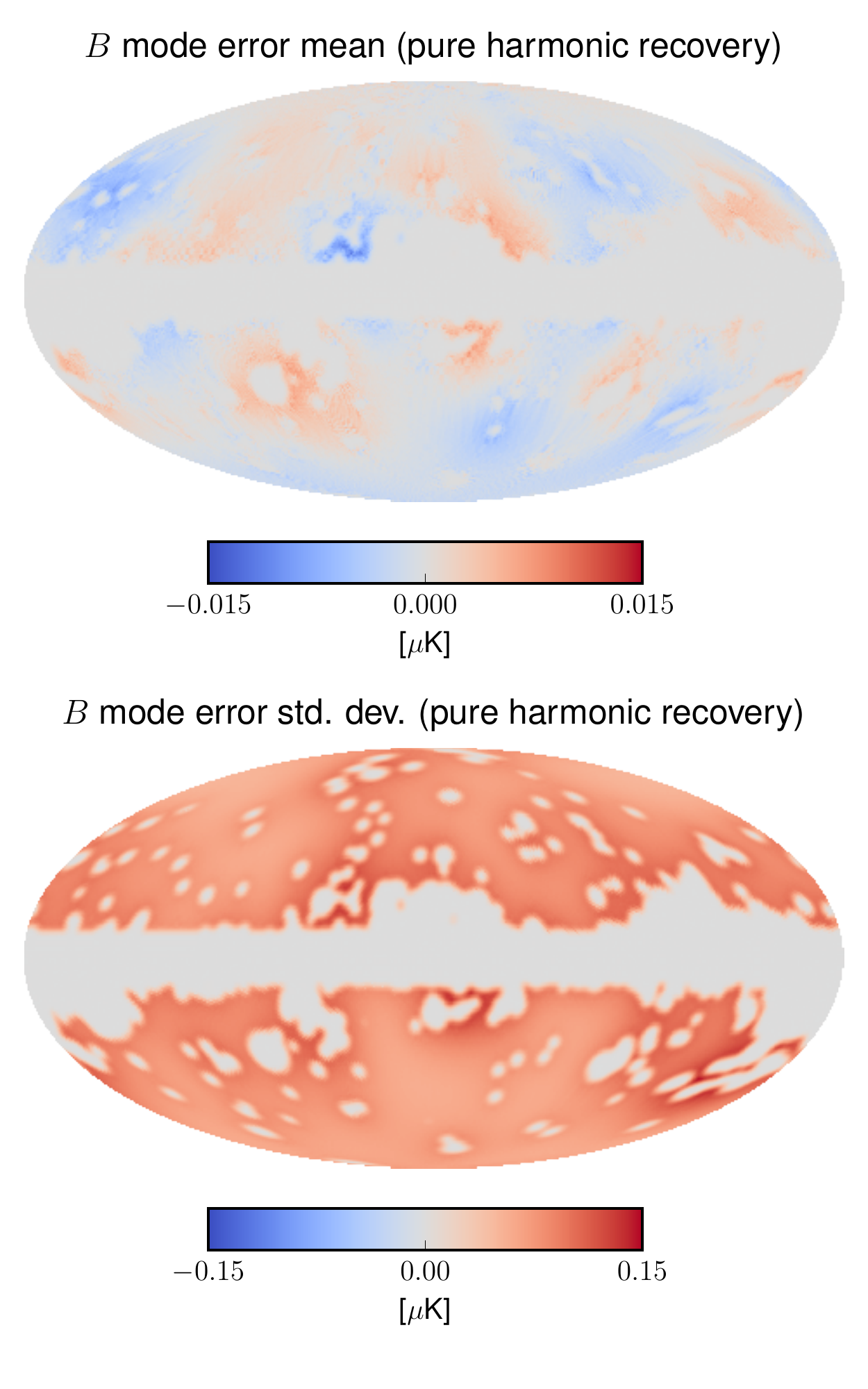}
\includegraphics[width=5.8cm]{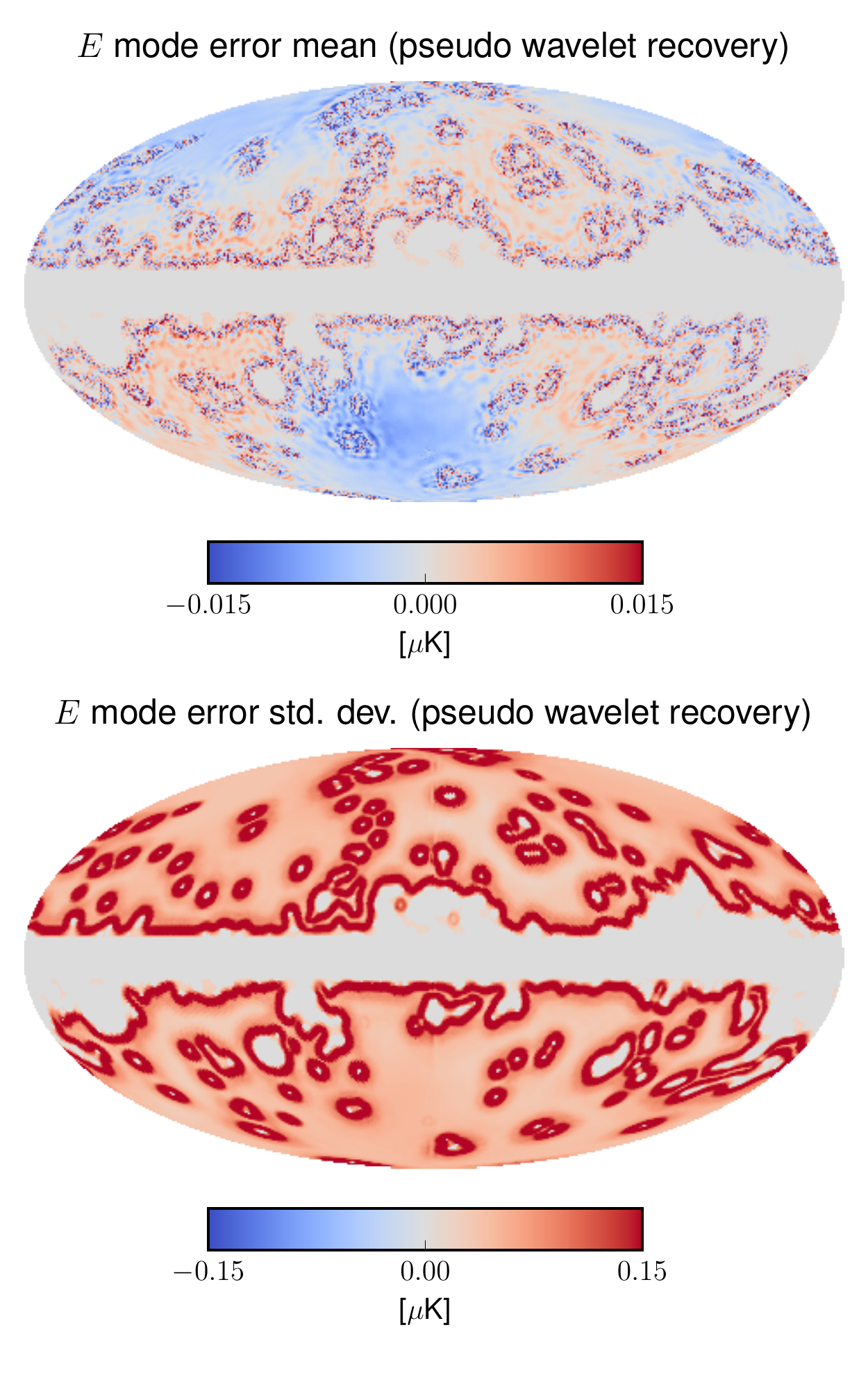}
\includegraphics[width=5.8cm]{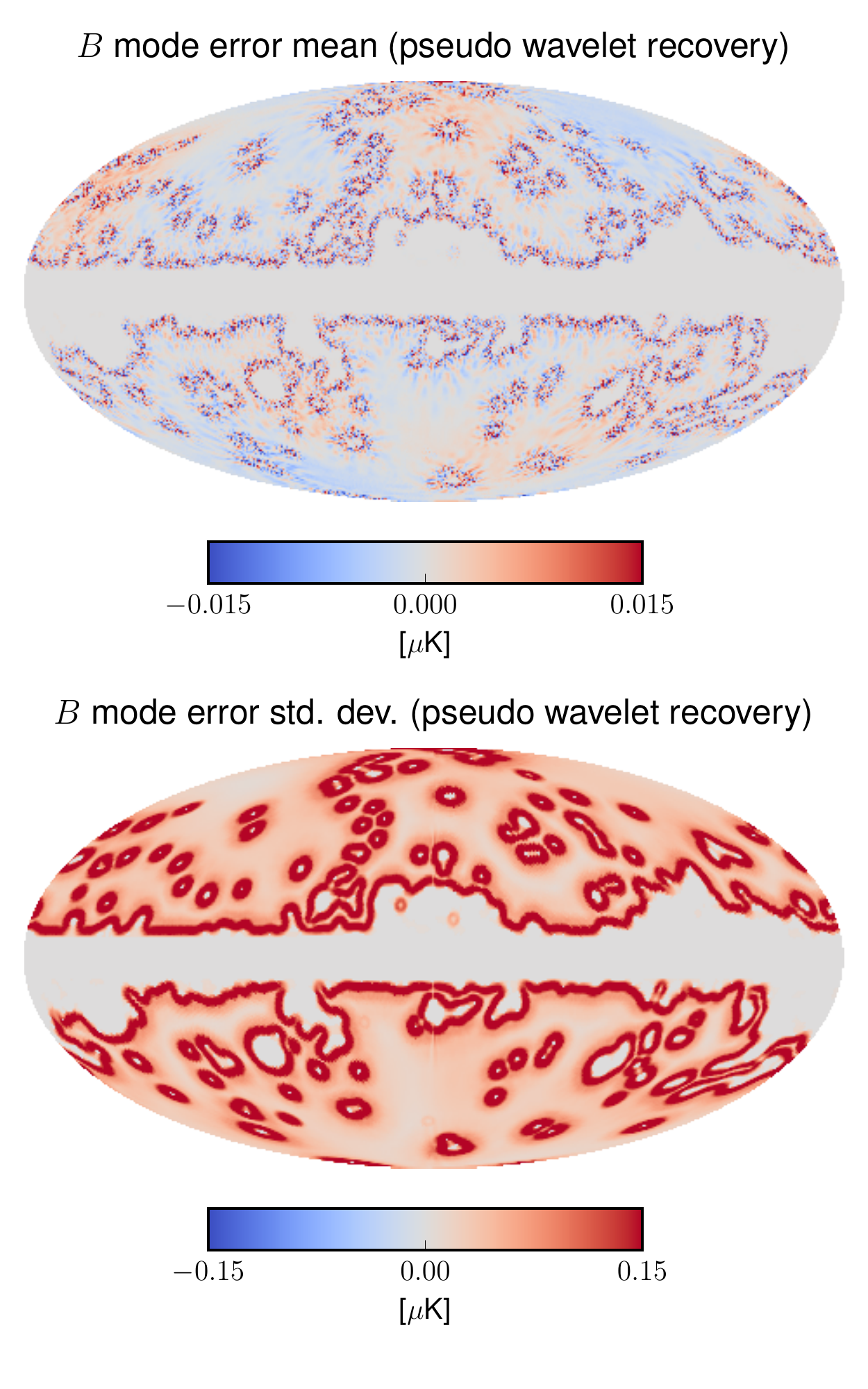}
\includegraphics[width=5.8cm]{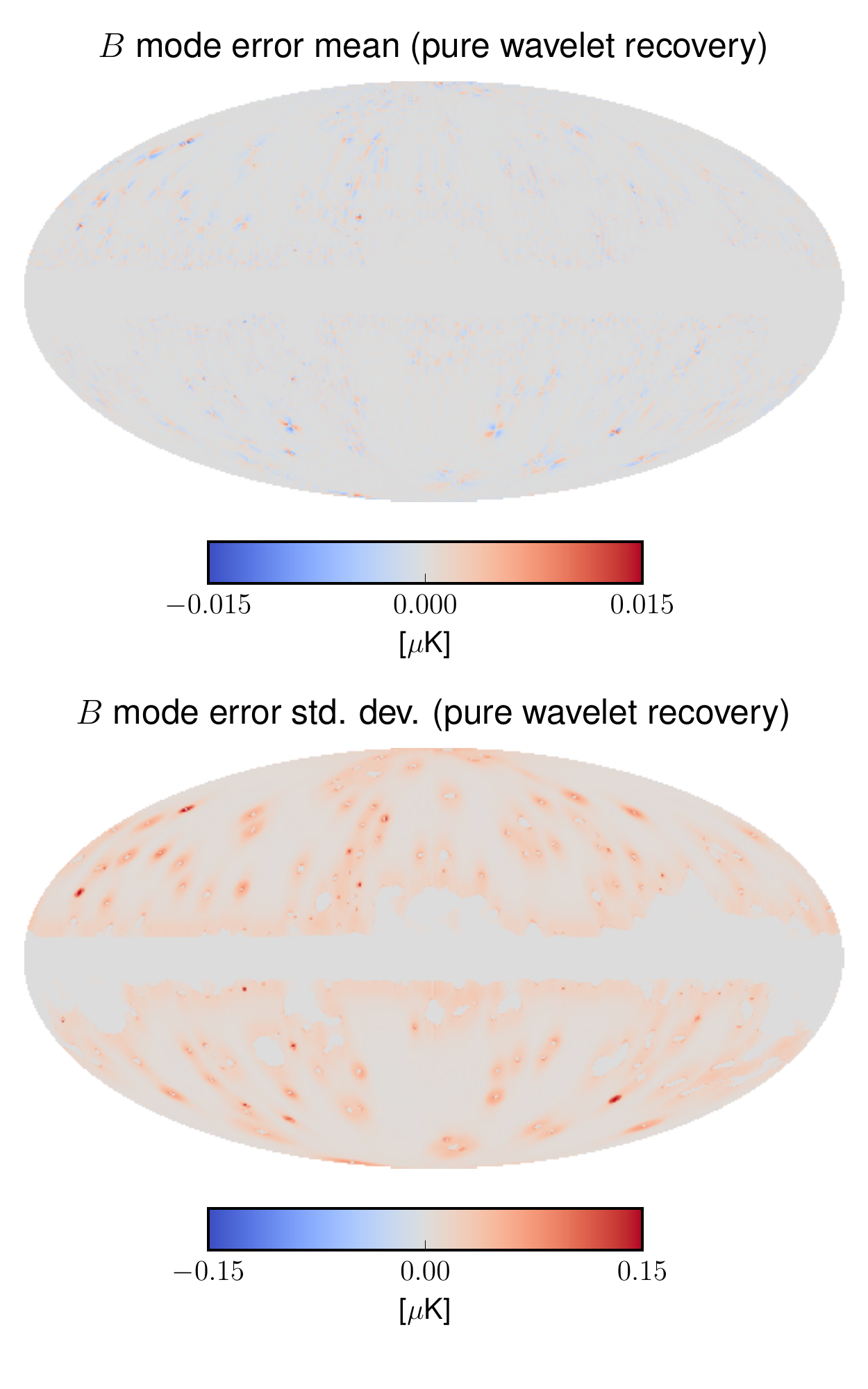}
\caption{Mean and standard deviation of the \E-\B error maps using the harmonic and wavelet estimators, without and with cancellation of the ambiguous modes (pseudo and pure recovery, respectively). 
With the standard (pseudo) recovery, the reconstructed \E-\B maps are unbiased in most of the sky, and small levels of leakage are present near the edges of the mask.
With the pure recovery, most of the small scale leakage is successfully cancelled, leading to significantly improved maps. Notice that the wavelet pure estimator is superior to the harmonic pure estimator, yielding percent-level residuals. 
The residual leakage on large scales in the case of the harmonic approach is due the inability to perform scale-dependent masking and, while the mask used is effective on small scales, it is not on large scales (see text for discussion).}
\label{fig:ebrecmaps}
\end{figure*}

In the previous section we introduced the pure estimator in wavelet space assuming a single mask $M(\ang)$, as in the harmonic approach.
However, this neglects a significant degree of freedom offered by the wavelet space approach: the possibility of exploiting a scale- and orientation-dependent masking scheme, \ie a sequence of masks $M^{j}(\rho)$ with $J_0\leq j \leq J$.  Apodised masks of different sizes can be constructed and applied to each wavelet scale, matching the size of the apodised mask to the scale of the modes probed by a given wavelet scale.  Furthermore, apodised masks can be matched to the directional structure of the original mask that is probed for each wavelet orientation.  Moreover, the scale- and orientation-dependence afforded by the wavelet space approach is exploited simultaneously.
The wavelet space approach naturally supports simultaneous localisation in space, scale and direction, which cannot be easily incorporated into existing pure harmonic reconstruction techniques.  We generalise the pseudo and pure wavelet estimators presented previously to scale- and orientation-dependent masks.  

Since the orientation of wavelet coefficients $\gamma$ is discretised as $\gamma_1,\ldots,\gamma_N$, we index scale- and orientation-dependent masks as $M^{jn}(\ang)=M^j(\phi,\theta,\gamma_n)$, noting that the Euler angles for a given $\gamma$ map onto the sphere as $(\alpha,\beta)=(\phi, \theta)$.  An additional mask is also required for the scaling function; as before we do not include the equations given the similarity with the case of wavelets.
In the case of scale- and orientation-dependent masking, our notation for the masks and their derivatives becomes
\eqn{
	{}_0 M^{jn} = M^{jn}, \quad {}_{\pm 1} M^{jn} = \spinpm M^{jn}, \quad {}_{\pm 2} M^{jn} = \spinpm^2 M^{jn},
}
and the products with the input Stokes parameters are denoted by
\eqn{
	{}_{\pm 2}\widetilde{P}^{jn} = {}_0 M^{jn} {}_{\pm 2}P, \quad {}_{\pm 1}\widetilde{P}^{jn} = {}_{\mp 1} M^{jn} {}_{\pm 2}P, \quad {}_{\pm 0}\widetilde{P}^{jn} = {}_{\mp 2} M^{jn} {}_{\pm 2}P,
}
where, again, we have omitted the dependency on $\ang$ for concision. 

The standard pseudo estimator in wavelet space then reads
\eqn{
  \widetilde{W}_{\epsilon}^{{}_0\Psi^j} (\rho) \hspace*{-1mm} &=&  - {\rm Re} \left[  {W}_{{}_{\pm 2} \widetilde{P}^{jn}}^{{}_{\pm2}\Psi^j} (\rho) \right] \\
  \widetilde{W}_{\beta}^{{}_0\Psi^j} (\rho) \hspace*{-1mm} &=&  \mp {\rm Im} \left[  {W}_{{}_{\pm 2} \widetilde{P}^{jn}}^{{}_{\pm2}\Psi^j} (\rho) \right],
}
while the pure estimator reads
\eqn{
	\widehat{W}_{\epsilon}^{{}_0\Psi^j}(\rho) &=&  -\ {\rm Re} \left[  W_{{}_{\pm2}\widetilde{P}^{jn}}^{{}_{\pm2}\Upsilon^{j}}(\rho) + 2 W _{{}_{\pm1}\widetilde{P}^{jn}}^{{}_{\pm1}\Upsilon^j}(\rho) + W_{{}_0\widetilde{P}^{jn}}^{{}_0\Upsilon^j}(\rho) \right]\\
	\widehat{W}_{\beta}^{{}_0\Psi^j}(\rho) &=&  \mp\ {\rm Im} \left[  W_{{}_{\pm2}\widetilde{P}^{jn}}^{{}_{\pm2}\Upsilon^j}(\rho) + 2 W_{{}_{\pm}1\widetilde{P}^{jn}}^{{}_{\pm1}\Upsilon^j}(\rho) + W_{{}_0\widetilde{P}^{jn}}^{{}_0\Upsilon^j}(\rho) \right] ,
}
where we have slightly abused the wavelet coefficient notation.  
In essence, masking of the Stokes parameters is performed in wavelet space, where the scale- and orientation-dependence of masks is accessible.  Once wavelet coefficients of the \E and \B mode estimators are computed by the above expressions, \E and \B mode maps can be computed by inverse scalar wavelet transforms.

Since multiple masks are involved, constructing a power spectrum estimator is complicated when considering scale- and orientation-dependent masking.
However, a good approximation can be written by using the results of the standard harmonic space spectrum estimators.
We start by representing the true, full sky power spectra in terms of the power spectra of the wavelets for various scales $j$ and orientations $n$:
\eqn{
	{C}_\ell^E = \sum_{jj'nn' \in \mathcal{S}(\ell)} {C}_\ell^{E,jnj'n'} \label{eq:wavpowspec0}\\
	{C}_\ell^B = \sum_{jj'nn' \in \mathcal{S}(\ell)} {C}_\ell^{B,jnj'n'}, \label{eq:wavpowspec1}
}
where $\mathcal{S}(\ell)$ represents the set of indices for which the wavelets have a non-zero contribution for a given $\ell$.
This decomposition highlights that the total power spectra are a sum of ${C}^{E,jnj'n'}$ and ${C}^{B,jnj'n'}$, which are cross-power spectra of the various wavelet scales and directions. 
A simple, approximate estimator for ${C}_\ell^E$ and ${C}_\ell^B$ is obtained by replacing ${C}^{E,jnj'n'}$ and ${C}^{B,jnj'n'}$ with estimates calculated via the standard pseudo or pure power spectrum estimators described in the previous section. 

\begin{figure*}
\hspace*{-7mm}\includegraphics[width=19cm]{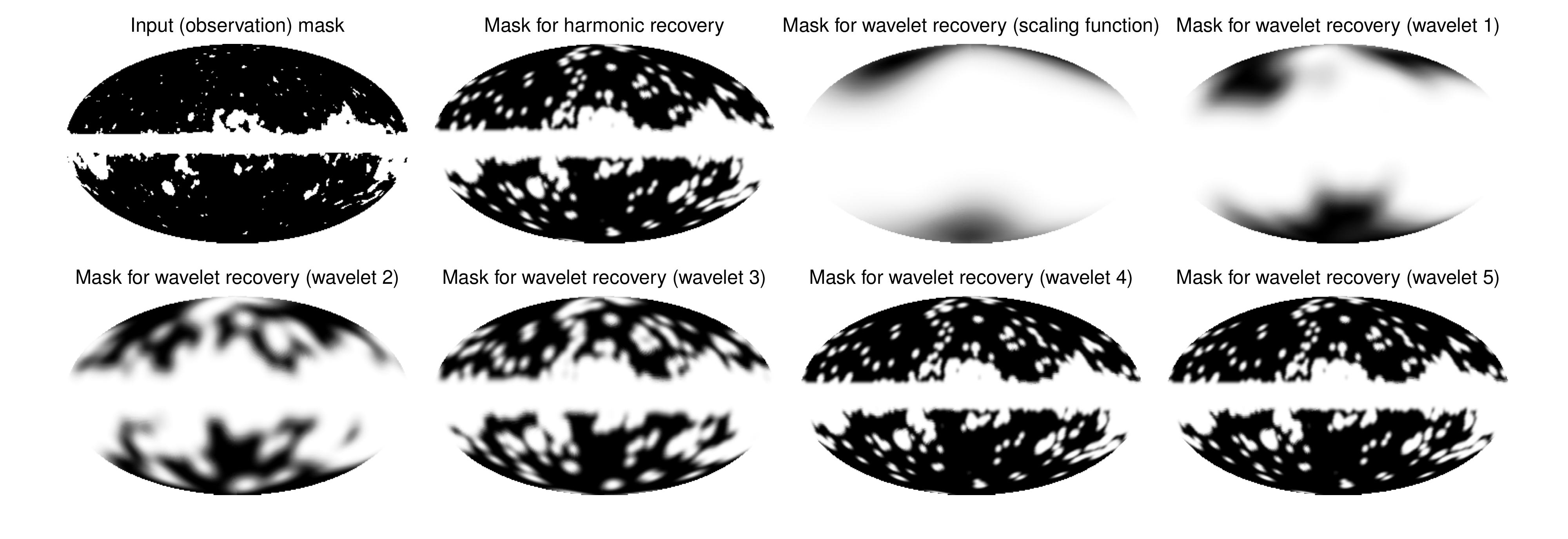}
\caption{Harmonic and wavelet processing masks applied to the Stokes $Q\pm i U$ maps before the \E-\B reconstruction. They are constructed by smoothing, thresholding, and smoothing the initial observation mask (top left panel) with a beam described in Appendix \ref{sec:beam}. The apodisation lengths, detailed in the text, are adjusted to the scale under consideration and allow for a highly accurate recovery of the \E and \B modes.}
\label{fig:masks}
\end{figure*}

\section{Illustration on simulations}\label{sec:results}

We apply the \E-\B reconstruction techniques outlined previously to simulations and evaluate their performance. In order to focus on the main novelty of this paper---the \E-\B reconstruction algorithm in wavelet space---and to illustrate its main features, we consider relatively simple simulations with moderate band-limits and a simple masking scheme.  Nevertheless, these simulations are adequate for studying and comparing the properties of the different reconstruction schemes, including the ability to deal with leakage more effectively and recover improved \E and \B maps.  We consider a single masking scheme for all methods (\ie consisting of particular choices for smoothing and for apodisation lengths). The advantages of the wavelet approach are independent of the masking scheme and details of the simulation. 
In particular, one could further optimise both approaches by creating optimal masks for the harmonic space as well as for the various wavelet scales involved without compromising the flexibility and advantages of the new approach.

All estimators (\ie all combinations of pseudo/pure and harmonic/wavelet \E-\B estimators) are implemented in the \ebsepcode\footnote{\url{http://www.ebsep.org}} code, which will be made public following further testing and evaluation.
\ebsepcode\ relies on the \stwoletcode\footnote{\url{http://www.s2let.org}} code
\citep{leistedt:s2let_axisym,McEwen:2015s2let} to perform scalar and spin wavelet transforms, which in turn relies on the 
\sshtcode\footnote{\url{http://www.spinsht.org}} code
\citep{mcewen:fssht} to compute spherical harmonic transforms, the
\sothreecode\footnote{\url{http://www.sothree.org}} code
\citep{2015ISPL...22.2425M} to compute Wigner transforms and the
\fftwcode\footnote{\url{http://www.fftw.org}} code to compute Fourier
transforms.  Note that \ebsepcode\ also supports the analysis of data on the
sphere defined in the common
\healpix\footnote{\url{http://healpix.jpl.nasa.gov}}
\citep{gorski:2005} format.

\begin{figure}
\hspace*{-5mm}\includegraphics[width=9.2cm]{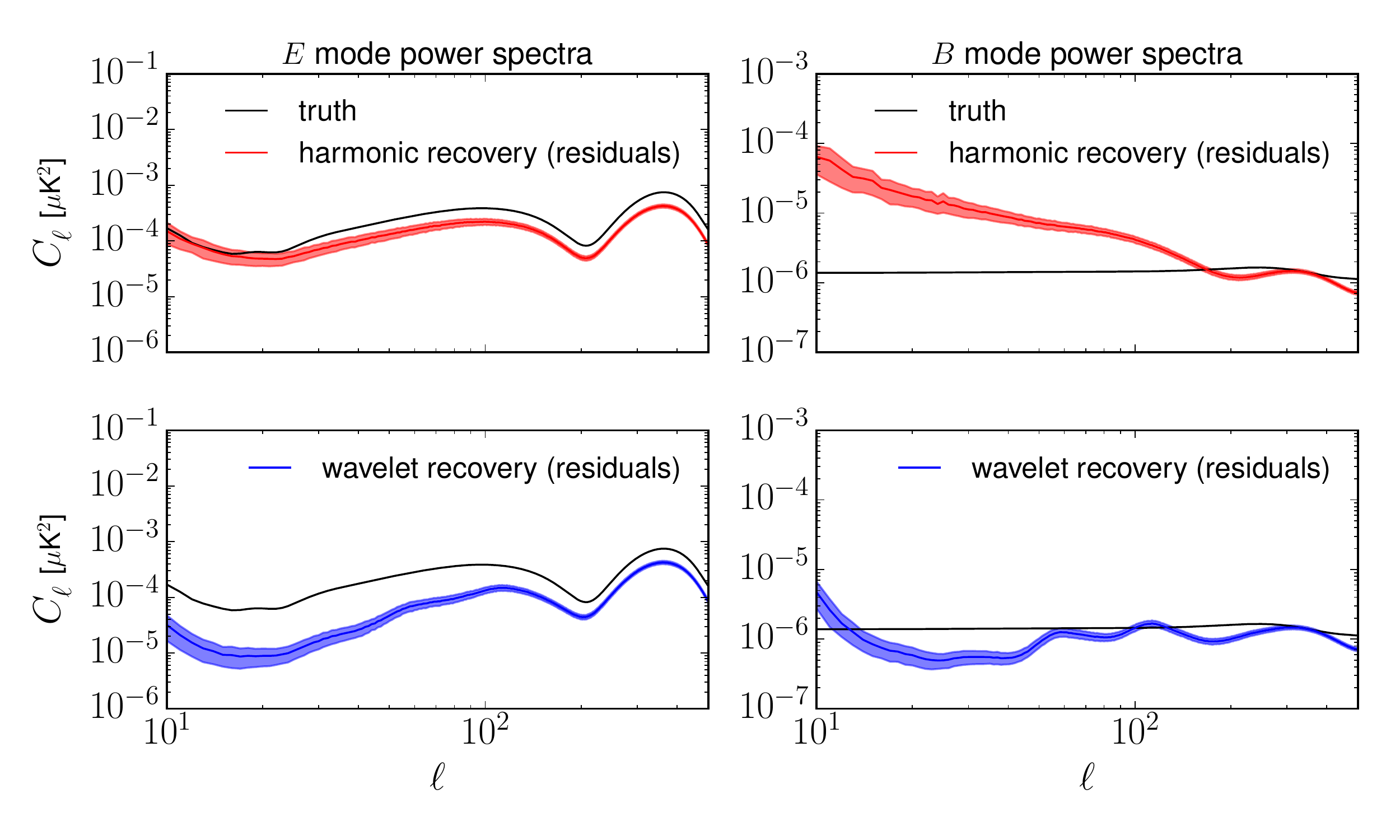}
\caption{Mean and standard deviation of full sky power spectra of residual \E-\B modes shown in \figref{fig:ebrecmaps}, using the pseudo harmonic and wavelet estimators, \ie without cancellation of the ambiguous modes (red/blue curve and shaded region). The input power spectra are also shown for comparison (solid black curve).  Residuals of the pseudo wavelet approach are reduced by approximately an order of magnitude compared to the pseudo harmonic approach.}
\label{fig:powspecnopure}
\end{figure}

\begin{figure}
\hspace*{-5mm}\includegraphics[width=9.2cm]{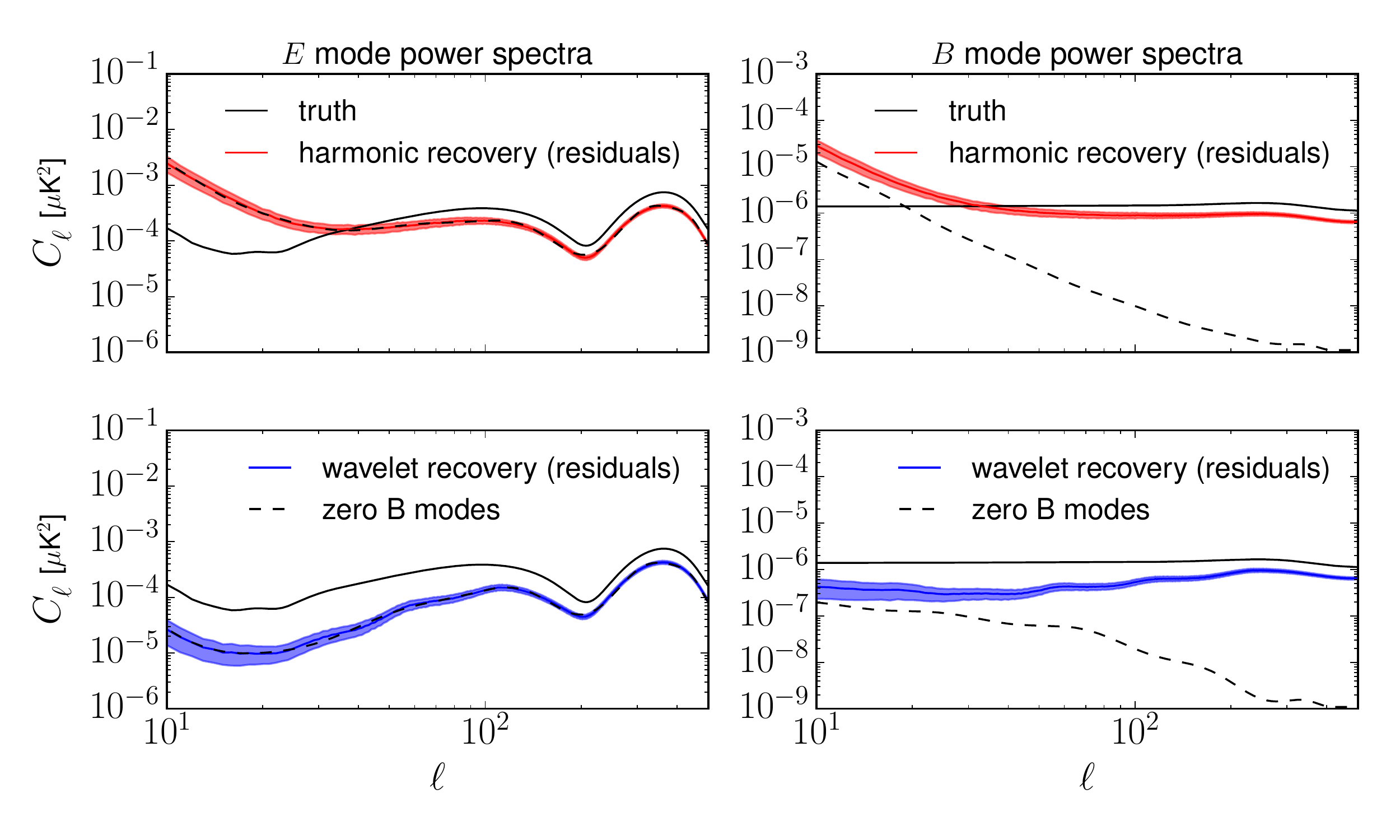}
\caption{Same as \figref{fig:powspecnopure} but using the pure harmonic and wavelet estimators, \ie with cancellation of the ambiguous modes. Residuals of the pure wavelet approach are reduced by approximately two orders of magnitude compared to the pseudo harmonic approach.  Some residual bias remains due to an imperfect cancellation of the leakage resulting from the mask not being optimised. {The dashed line shows the same results but for a zero input $BB$ power spectrum, to highlight the fraction of residuals due to \E-to-\B leakage and \B-to-\B mixing.} All masks in the harmonic and wavelet approaches could be optimised without altering the conclusion that the scale-dependent masking of the wavelet approach makes it superior to the standard harmonic method.}
\label{fig:powspecpure}
\end{figure}

\subsection{Parameters and masks}

For these tests we consider the band-limit $L=512$, which is sufficient to highlight the effectiveness of the \E-\B reconstruction while keeping the computation tractable without resorting to high-performance computing infrastructure. 
Indeed, while our wavelet transform has been tested up to $L=4096$ and runs in minutes for moderate resolutions (scaling as $JNL^3$), it must be run several times on each simulation in order to produce the maps and power spectra of interest (see details below).
For the wavelet transforms, we use $\lambda=2$, $J_0=5$ and $N=1$, yielding $J=9$, and therefore five wavelet scales and one scaling function. 

We simulate 1000 noiseless \Q and \U maps from \E and \B power spectra.
These were obtained with {\sc camb}\footnote{\url{http://camb.info}} for the flat $\Lambda$CDM model with the following cosmological parameters: $H_0=67.5$, $\Omega_bh^2=0.022$, $\Omega_ch^2=0.122$, $m_{\nu}=0.06$, $\Omega_k=0$, $\tau=0.06$, $n_s=0.965$. 
We set the tensor-to-scalar ratio to $r=0$ but use lensed power spectra, such that the information contained in the $B$ modes is due to lensing of the $E$ modes.

We use the binary {\it Planck} UB77 mask,\footnote{\url{http://irsa.ipac.caltech.edu/data/Planck/release_2/ancillary-data}} shown in the upper left panel of \figref{fig:masks}, and referred to as the `observation mask'. 
The `processing mask' involved in the harmonic space recovery is shown in the upper center panel of the same figure. 
It is obtained by smoothing the observation mask, thresholding it at a level of $0.99$ (all pixels with a value lower that $0.99$ are set to $0$, and the others are set to $1$), and smoothed again. 
For both smoothing operations we use the beam presented in Appendix~\ref{sec:beam} with an apodisation length of $4\pi/L$, \ie smoothed with $b_{\ell}(4\pi/L)$, where $b_\ell$ are the harmonic coefficients of the beam. 
The processing masks for the scaling function and wavelet coefficients are constructed in the same manner; the former is smoothed with $b_{\ell}(4\pi/\lambda^{J_0-1})$, and the latter with $b_{\ell}(4\pi/\lambda^{j})$ for $j=J_0,\ldots,J$.
All processing masks are shown in \figref{fig:masks}.

\subsection{Results}

We now assess the quality of the reconstructed \E-\B maps for both the harmonic and wavelet reconstructions, with and without ambiguous mode cancellation (\ie both the pure and pseudo estimators, respectively). 
We focus on quantifying the mean and standard deviation of the residuals in the \E-\B maps (which is mostly due to \E-\B mixing). 
For each simulation, we reconstruct \E and \B maps and compare them to (\ie subtract) the true simulated input maps. 
Because this comparison is only valid in the unmasked region, the true \E-\B maps have to be appropriately masked. 
In the case of the harmonic approach, this simply requires the application of the processing mask, while for wavelet reconstruction the wavelet coefficients of the true \E-\B maps must be masked using multiple processing masks (via wavelet space).

The residual \E and \B modes (\ie the mean and standard deviation of the difference between the reconstructed and true maps) for the pseudo estimation methods (with no cancellation of the ambiguous modes) in harmonic and wavelet spaces are shown in \figref{fig:ebrecmaps} (first and second column).
The colour scales of the residuals are scaled to one tenth of the standard deviation of the true \E-\B maps, which is roughly the amplitude of the expected leakage near the edges of the mask.
For both reconstruction methods the reconstructed \E-\B maps are unbiased on most of the sky, and a small residual bias occurs near the edges of the mask.
The leakage for the pseudo wavelet estimator is smaller than for the pseudo harmonic estimator, as a consequence of exploiting scale-dependent masking in the wavelet approach by adjusting the apodisation length for each wavelet scale separately.  
We do not exploit directionality in the simple simulations presented here and consider axisymmetric wavelets only.

The bias and standard deviation of the residual \E-\B modes in the case of the pure estimator (\ie with cancellation of the ambiguous modes) are also shown in \figref{fig:ebrecmaps} (third column).
We do not show the \E-mode maps in the case of the pure estimator since this is typically of lower interest given that \B-to-\E leakage is significantly smaller than \E-to-\B leakage in the case of the CMB. 
For both the harmonic and wavelet space reconstructions, the small scale leakage which was prominent in the standard reconstruction is almost entirely cancelled.
The cancellation of the large scale leakage is also effective in the case of the wavelet reconstruction.
However, the harmonic space reconstruction suffers from some residual large-scale leakage in both the \E and \B modes. 
This is due to the imperfect mask being used. 
Indeed it is not possible to exploit scale-dependent masks for the harmonic estimators and only a single mask can be applied. In particular, the apodisation length is too small for the ambiguous mode cancellation to be effective on large scales. 
This large-scale leakage could be mitigated by employing a significantly apodised mask but, as a consequence, small-scale leakage would not be removed as effectively.
The wavelet approach, however, provides a natural solution to this issue: by adapting the apodisation length to each scale, the ambigious modes of different scales are removed effectively.
The scale-dependent masking of the wavelet approach with a simple smoothing kernel is sufficient to cancel out most of the leakage and recover high quality \E-\B maps. 
This is similar to state-of-the-art \E-\B power spectrum estimators which employ the harmonic method in separate multipole bins with optimised masks.
In the current work, overlapping multipole bins are used, and the reconstruction is formulated in a wavelet basis, where the notions of scale and direction are well-defined.

Angular power spectra of {the full sky} residuals maps (\ie reconstructed minus true \E-\B maps) are shown in \figref{fig:powspecnopure} for the standard, pseudo estimators (\ie with no ambiguous mode cancellation). 
We measure full sky angular power spectra without correcting for the residual leakage in order to highlight its amplitude and the scales affected.
The true power spectra used to simulate the maps are also shown for comparison.
As expected, the \E modes are mostly perfectly recovered on all scales, but significantly contaminate the \B modes. 
With the standard harmonic reconstruction (using a single mask), the \E-to-\B leakage washes out the cosmological signal in the \B modes on large to intermediate scales ($\ell \leq 100$). 
Accounting for the residual mixing in the power spectrum estimator, for example via Eq.~\eqref{eq:couplingpseudo}, will lead to \B mode estimates that are unbiased but suffer from a large variance. 
Both the amplitude of the residuals and the variance of the power spectrum in the case of the wavelet reconstruction are smaller due to the use of scale-dependent masks, as also observed at the map level.
Reducing both the residuals and the variance of the estimates require cancellation of the ambiguous modes.

Angular power spectra of {the full sky} residual maps for the pure mode approach are shown in \figref{fig:powspecpure} (\ie with ambiguous mode cancellation). 
The residuals are significantly reduced, for both the harmonic and wavelet reconstructions.
However, the harmonic case still suffers from \E-\B mixing on large angular scales. 
As discussed above, this is due to the use of a single mask, yielding imperfect ambiguous mode cancellation.
The power spectrum estimator presented in Eq.~\eqref{eq:couplingpure} can account for this residual mixing, at the cost of a mild increase of the variance.
The wavelet estimator naturally addresses these issues, since it employs multiple masks and yields a more accurate cancellation of the ambiguous modes on all scales, with a smaller variance increase caused by \E-\B mixing.  
The residual $B$-mode is reduced by approximately two orders of magnitude for the pure wavelet estimator (\figref{fig:powspecpure}) when compared to the pseudo harmonic estimator (\figref{fig:powspecnopure}).
{Note that this $B$-mode residual is due to both \E-\B leakage and \B-\B mixing resulting from the non-trivial effect of ambiguous-mode cancellation; this could be corrected by deconvolving the results into full-sky power spectra with the estimators detailed in \cite{Grain2012crosspure, Ferte2014pureEB} for instance.}

\section{Discussion and future perspectives}\label{sec:discussion}

We presented a new formalism for extracting \E and \B modes from spin 2 signals via a spin wavelet transform, and validated it on simple simulations. 
We anticipate this new approach will be useful for analysing CMB polarisation and cosmic shear data. 
Most studies of these observables focus on compressing the information into \E and \B power spectra in order to efficiently constrain cosmological parameters.
However, studying maps of the \E and \B modes offers complementary insights into cosmological models and also new angles for searching for exotic physics. 
For instance, patterns in the CMB \E-\B modes around features in the temperature field (\eg hot and cold spots) are not only useful cosmological probes but also powerful checks of the spatial quality of the data. 
They can be used to test for new physics such as non-Gaussianity, phase transitions or exotic topology imprinted in the CMB  \citep[\eg][]{barreiro:1997, hobson:1999, durrer:1999, barreiro:2000, cayon:2000, bh:2001, davis:2005, cruz:2006a, gonzalez:2006, mcewen:2007:isw3, bridges:2007:bianchi, feeney:2011a, feeney:2011b}.
Similarly, patterns in the cosmic shear \E-\B modes around massive structures and voids are sensitive probes of cosmological parameters \citep[see \eg][ for recent studies]{Kacprzak:2016vir, Pujol:2016lfe, Gruen:2015jhr}.
In both cases wavelets have proved to be a powerful basis for extracting such features (see references above).

One of the core advantages of a spin wavelet approach to \E-\B separation is the scale- and orientation-dependent weighting and masking that it affords due to the spatial, scale and directional localisation properties of the scale-discretised wavelets adopted \citep{McEwen:2015s2let,mcewen:2015s2let_localisation}.  Different apodised masks can be constructed and applied to each wavelet scale and orientation, matching the size and shape of the apodised mask to the structure probed by a given wavelet.

Nevertheless, we adopt a simple scheme for constructing apodised masks and have not yet investigated the construction of optimal masks that minimise leakage and yield optimal, leakage-free \E-\B maps.  Such optimal masks could be constructed in an analogous manner to the optimal masks used to obtain minimum variance \E-\B binned power spectra \cite[\eg][]{Smith2006pureEB, SmithZaldarriaga2007pureEB,  Grain2012crosspure, Ferte2014pureEB} and is a avenue of future research.  Furthermore, although we present the general directional wavelet formalism for \E-\B separation, we have not yet applied directional wavelets (with $N>1$) to exploit the orientation-dependence that the formalism provides, which is also left for future work.

Another key advantage of the \E-\B reconstruction method presented here is the ability to naturally interface in a coherent and efficient manner with other pre- or post-processing (\ie using \Q-\U or \E-\B, respectively) algorithms that exploit wavelets.
A concrete example is the blind extraction of the CMB signal from multi-frequency microwave observations using an internal linear combination (ILC) method \citep[\eg][which can also be used to extract the thermal Sunyaev-Zel'dovich signal]{2004ApJ...612..633E, 2005EJASP2005..100M, 2008StMet...5..307B, 2009A&A...493..835D, 2012MNRAS.419.1163B, 2013MNRAS.435...18B}.
ILC approaches have proved successful for creating clean CMB maps with no or weak assumptions about the foregrounds. 
While good models are now available for temperature CMB foregrounds, this is not the case for CMB polarisation. 
\cite{Rogers:2016silc} presented SILC, the first ILC method exploiting directional wavelets, which proved useful to improve the quality of the reconstruction of the CMB temperature on small scales.
This approach was extended in \cite{Rogers:2016spinsilc}, which presented Spin-SILC, the first ILC method exploiting spin directional wavelets \citep{McEwen:2015s2let} to analyse CMB polarisation. 
Spin-SILC is therefore able to exploit the spin nature of the data, simultaneously producing clean \Q, \U, \E and \B maps via the properties of spin wavelets presented in the present paper.
Future applications of the Spin-SILC algorithm of \cite{Rogers:2016spinsilc} will involve cut-sky data and deal with \E-\B leakage using the ambiguous mode cancellation in wavelet space developed here. 
Post-processing methods such as spatially-localised feature extraction could also be directly interfaced with this \Q-\U-to-\E-\B method.

As highlighted before, cosmic shear is also a spin 2 signal, so all examples given above for the CMB also apply to it.
In particular, the wavelet \E-\B reconstruction method allows one to directly analyse $\gamma$ maps and study the lensing of massive clusters, shear peaks, and voids.  
However, a major difference with the CMB is the possibility of studying cosmic shear in three dimensions by adding redshift as a radial dimension \citep{heavens:2003, castro:2005, kitching:2014}.
This is also supported in our framework, and can be achieved by adopting the spin 3D wavelets presented in \cite{Leistedt:2015vwa} (initially introduced in \cite{leistedt:flaglets} for the scalar setting).

In summary, we have developed new wavelet space approaches to \E-\B separation, including both pseudo and pure estimators.  We have demonstrated the validity of these new methods on relatively simple simulations, highlighting differences with the existing harmonic space method and showing how scale-dependent masking and ambiguous mode cancellation can be achieved. 
In future work we will study more realistic simulations, exploit directionality, optimise wavelet parameters (similar to that done in \citealt{Rogers:2016silc, Rogers:2016spinsilc}), and optimise masks.  Nevertheless, these extensions will not change the main features of our wavelet approach to \E-\B separation or our main conclusions.  The wavelet pseudo and pure \E-\B estimators developed here are highly effective for recovering \E and \B mode maps for CMB polarisation and cosmic shear, which in turn are important for going beyond power spectra analyses of cosmological data.

\section*{Acknowledgements}
\label{sec:ack}

We thank Keir K. Rogers and Andrew Pontzen for valuable discussions. HVP and BL were partially supported by the European Research Council under the European Community's Seventh Framework Programme (FP7/2007-2013) / ERC grant agreement number 306478-CosmicDawn. JDM was partially supported by the Engineering and Physical Sciences Research Council (grant number EP/M011852/1).

\appendix

\section{Compact smoothing beam}\label{sec:beam}

We seek to construct a function $h(\theta, \phi)$ (referred to as smoothing or apodising beam below) that is localised and smooth in real space such that a binary mask on the sphere convolved with this beam will satisfy the Dirichlet and Neumann boundary conditions.
In other words, the smoothed mask and its derivative must vanish at the boundaries of the (un-smoothed) mask. 

We restrict our interest to axisymmetric beams $h(\theta, \phi) = h(\theta,0), \ \forall \phi$. In this case the spherical harmonic coefficients of the beam are such that $h_{\ell m}=0,\ \forall m\neq0$, so we will use the compressed notation $b_\ell = h_{\ell 0}$. 
Performing the smoothing in pixel space via a convolution is possible (especially since the beam is localised) but typically inaccurate. 
Fortunately, convolution of a mask $M(\ang)$ with axisymmetric functions can be computed efficiently in harmonic space by multiplying the spherical harmonic coefficients of the mask $M_{\ell m}$ with $\sqrt{\frac{4\pi}{2\ell+1}} b_\ell$. 
Hence, our goal is to construct a beam $b_\ell$ with a well-localised real space representation. 

It was shown in \cite{SmithZaldarriaga2007pureEB} that the optimal apodisation beam for power spectra proportional to $\ell^4$ (which is approximately the case for the $\epsilon$ and $\beta$ fields) in the Euclidean, one dimensional region $r\in[0,R]$ reads
\eqn{
	W_{\rm flat}(r,R)  = 1 - \frac{\ell_+ \ I_1(\ell_+R) \ I_0(\ell_-r) \ - \ \ell_- \ I_1(\ell_-R) \ I_0(\ell_+r)}{\ell_+ \ I_1(\ell_+R) \ I_0(\ell_-R) \ - \ \ell_- \ I_1(\ell_-R) \ I_0(\ell_+R)},
}
where $r$ is the distance to the boundary and $R$ is the apodising length, or size of the domain to be smoothed. $I_0$ and $I_1$ are modified Bessel functions, and we have set
\eqn{
	\ell_{\pm} = \ell_0 \sqrt{2\pm\sqrt{3}}.
	}
$\ell_0$ is the multipole under consideration. 
For our tests, we set $\ell_0$ to $\frac{\pi}{R}$; this is somewhat more restrictive than the kernel of \cite{SmithZaldarriaga2007pureEB} but greatly simplifies the construction of the beam since $R$ is now the only parameter and corresponds to the apodising length.  
Keeping $\ell_0$ as a parameter would greatly complicate the task of designing kernels for the various wavelet scales involved in the \E-\B reconstruction.
Exploiting more flexible (and optimal) smoothing kernels is left for future work since it does not affect the demonstration of the methods presented here.

The Euclidean setting above can be extended to the sphere by identifying $r$ with the colatitude $\theta$, \ie the distance from the North pole. 
In this case, $ h(\theta, \phi) = W_{\rm flat}(\theta)$ is a valid axisymmetric smoothing function.
The spherical harmonic coefficients for $m=0$ reduce to
\eqn{
	W_\ell(R) =\int_0^\pi  W_{\rm flat}(\theta,R)P_\ell(\cos\theta)  \sin \theta\ {\rm d}\theta.
}
Thanks to the analytical expressions that exist for the integration of Bessel functions and Legendre polynomials, we find 
\eqn{
	W_\ell(R) &=& \frac{R^2}{2} \ {}_0F_1\Bigl(2, -\frac{R^2\ell^2}{4}\Bigr) \\
	 - &&\hspace*{-4mm}  \frac{
		\ell_+ \ I_1(\ell_+R) \ C(R,\ell_-,\ell+\half) \ - \ \ell_- \ I_1(\ell_-R) \ C(R,\ell_+,\ell+\half)
	}{
		\ell_+ \ I_1(\ell_+R) \ I_0(\ell_-R) \ - \ \ell_- \ I_1(\ell_-R) \ I_0(\ell_+R) \nonumber
	}
}
with the function
\eqn{
	C(R,\ell_\pm,\ell) = \frac{R}{\ell_\pm^2+\ell^2} \Bigl(\ell_\pm \ I_1(\ell_\pm R) \ J_0(\ell R) + \ell \ I_0(\ell_\pm R) \ J_1(\ell R) \Bigr),
}
where ${}_0F_1$ is the confluent hypergeometric function and $J_0$ and $J_1$ are Bessel functions.
This new result provides an accurate method to compute the smoothing beam analytically in harmonic space, adapting to the sphere the kernel of \cite{SmithZaldarriaga2007pureEB} (optimal for $\ell^4$ power spectra).

For the tests described in this paper we use the normalised beam $b_{\ell}(R) = {W_{\ell}(R)}/{W_{0}(R) }$, with $R$ the apodising length under consideration.


\providecommand{\eprint}[1]{\href{http://arxiv.org/abs/#1}{arXiv:#1}}
\bibliographystyle{mymnras_eprint}	
\bibliography{bib}

\end{document}